\documentclass[letterpaper,twocolumn,10pt]{article}
\usepackage{usenix2019_v3}

\usepackage{tikz}
\usepackage{amsmath}
\usepackage[english]{babel}
\usepackage{blindtext}
\usepackage{enumitem}
\usepackage{titlesec}
\hypersetup{colorlinks=true,linkcolor=black}
\newenvironment{packed_itemize}{
    \begin{list}{\labelitemi}{\leftmargin=1.0em}
     \setlength{\itemsep}{2.5pt}
     \setlength{\parskip}{0pt}
     \setlength{\parsep}{0pt}
     \setlength{\headsep}{0pt}
     \setlength{\topskip}{0pt}
     \setlength{\topmargin}{0pt}
     \setlength{\topsep}{0pt}
     \setlength{\partopsep}{0pt}
    }{\end{list}}

\usepackage{filecontents}
\usepackage{caption}
\usepackage{subcaption}
\usepackage{bbding}
\usepackage{multirow}
\usepackage{listings}
\usepackage{tcolorbox}
\usepackage{algorithm}
\usepackage{algorithmic}
\usepackage{makecell}
\usepackage{booktabs}
\usepackage{pifont}

\newcommand{\edit}[1]{#1}
\newcommand{\camera}[1]{#1}
\newcommand{\chang}[1]{#1}

\long\def\com#1{}

\newcommand{\dlt}{LMT\xspace}
\newcommand{\dlts}{LMTs\xspace}
\newcommand{\pattern}{runtime behavior pattern\xspace}
\newcommand{\patternss}{runtime behavior patterns\xspace}
\newcommand{\cluster}{GPU cluster\xspace}
\newcommand{\clusters}{GPU clusters\xspace}
\newcommand{\sys}{\textsc{Eroica}\xspace}
\newcommand{\eg}{\textit{e.g.}\xspace}
\newcommand{\ie}{\textit{i.e.}\xspace}
\newcommand{\etc}{\textit{etc.}\xspace}

\newcommand{\para}[1]{\smallskip\noindent {\bf #1} }

\newcommand{\squishlist}{
   \begin{list}{$\bullet$}
    { \setlength{\itemsep}{0pt}      \setlength{\parsep}{3pt}
      \setlength{\topsep}{3pt}       \setlength{\partopsep}{0pt}
      \setlength{\leftmargin}{3.5mm} \setlength{\labelwidth}{1em}
      \setlength{\labelsep}{0.5em} } }

\newcommand{\squishend}{
    \end{list}  }

\usepackage{titling}
\postauthor{\end{tabular}\vspace{-0.7in}\end{center}}
\date{}

\usepackage{titlesec}
\titleformat{\subsection}
  {\normalfont\fontsize{11}{10}\bfseries}{\thesubsection}{1em}{}

\begin{document}

\date{}

\title{\Large {\bf \sys\hspace{-0.3em}\thanks{The title \sys shares its name with Beethoven’s Symphony No.~3 (Op.~55), a pinnacle of Classical music and a starting point of Romanticism, as \sys highly optimizes classic monitoring and profiling for \dlt performance troubleshooting while making a first attempt toward future AIOps.}\hspace{0.3em}: Online Performance Troubleshooting for Large-Scale Model Training} \vspace{-15pt}}

\author{
Yu Guan$^{1}$, Zhiyu Yin$^{1}$, Haoyu Chen$^{1}$, Sheng Cheng$^{1}$, Chaojie Yang$^{1}$, Kun Qian$^{1}$, Tianyin Xu$^{2}$,\\
Pengcheng Zhang$^{1}$, Yang Zhang$^{1}$, Hanyu Zhao$^{1}$, Yong Li$^{1}$, Dennis Cai$^{1}$, Ennan Zhai$^{1}$\\
{\it $^{1}$Alibaba Cloud, $^{2}$University of Illinois Urbana-Champaign}
\vspace{20pt}
}

\maketitle

\begin{abstract}
Troubleshooting performance problems of large model training (\dlt) is immensely challenging, due to unprecedented scales of modern \clusters, the complexity of software-hardware interactions, and the data intensity of the training process. Existing troubleshooting approaches designed for traditional distributed systems or datacenter networks fall short and can hardly apply to real-world training systems. In this paper, we present \sys, the first online troubleshooting system \chang{that provides both} \camera{fine-grained observation based on profiling, and} \chang{coverage of all machines in GPU clusters,} \camera{to diagnose performance issues in production, including both hardware and software problems (or the mixture of both).}
\sys{} effectively summarizes runtime behavior patterns of \dlt function executions via online profiling, and leverages differential observability to localize the root cause with minimal production impact. 
\sys has been deployed as a production service for large-scale \clusters of \camera{$\sim$100,000} GPUs for 1.5 years. It has diagnosed a variety of difficult performance issues with 97.5\% success.
\end{abstract}

\thispagestyle{empty}

\vspace{-5pt}
\section{Introduction}
\label{sec-intro}
\vspace{-5pt}

Troubleshooting performance problems of Large Model Training (\dlt) has become a grand challenge driven by unprecedented increases in AI model sizes and GPU-based training infrastructures. \dlt undertakes complex hardware-software interactions; unexpected behavior of hardware components (\eg, GPUs and their interconnects) to software code and configurations (\eg, data loaders, communication, and PyTorch), could lead to performance issues.
With rapid scaling of \dlt,
occurrences of performance problems inevitably increase accordingly, resulting in significant wastage of compute resources and production incidents~\cite{jiang2024megascale, HPN, SuperBench}. 

\com{Unfortunately, traditional troubleshooting techniques, designed for distributed systems~\cite{zhao2014lprof,barham2004using,terai2014extending,Network_Murphy_SIGCOMM2023} and data center networks~\cite{R_PingMesh, Host_NetBouncer,Host_007,Host_Passive}, fall short and can hardly help today's model training systems.
First, all existing approaches focus on CPU-based systems and 
    do not work with GPU-centric, heterogeneous systems.
Second, modern \dlt systems are orders of magnitude larger in parallel computation than traditional systems and networks.
The parallelism impairs existing approaches that
    rely on fine-grained traces and profiling data.
For example, an \dlt of 10,000 GPUs typically makes $4 \times 10^9$ function executions per second, which produces terabytes of profiling data.
Active measurement (\eg, probing system components~\cite{R_PingMesh,Host_NetBouncer,jha2020sc}) is not an option due to interference with production workloads.
Moreover, most existing approaches focus only on software defects or only on hardware; they cannot reason about their complex interactions.}

In practice, performance troubleshooting of production \dlt relies on data from either {\it online monitoring} or {\it offline profiling}.
\camera{Unfortunately, neither is ideal, and engineers often face difficult trade-offs between observation granularity and} \chang{cluster-wide coverage.
Online monitoring must be coarse-grained to cover the entire GPU cluster, including all \dlt training processes (\ie, workers) in all machines, without impacting performance. Offline profiling provides fine-grained observation but incurs substantial overhead and generates a huge amount of data that cannot be consumed in real time, so it cannot be enabled in real large-scale training; instead, engineers usually create a smaller test training environment for profiling, which lacks timeliness and is not always possible.
}

Most online monitors use vendor tools to collect performance counters of hardware components (\eg, GPUs~\cite{dcgm}, CPU~\cite{pcm}, CPU-GPU~\cite{dynolog}, and NICs~\cite{pcm,mstflint,R_PingMesh}) during \dlt, or collect timestamps of specific user function executions~\cite{ncclprofiler, ebpf}.
To avoid production impacts, typically the hardware sampling is only in second granularity, and only a few important user functions can be selected for monitor.
However, in our production experience, the coarse-grained observability can only detect general symptoms (\eg, performance degradation in forward/backward phases or in communication modules), but cannot pinpoint fine-grained root causes (\eg, which line of code or which network link caused the degradation).

Offline profiling (\eg, Torch Profiler~\cite{torch_profiler}, Nsight System~\cite{nsys}, and Nsight Compute~\cite{nsight_compute}) records all function executions (\eg, Python call stacks, CUDA executions) in \dlt and high-frequency hardware sampling (\eg, at 200k~Hz).
Based on these profiling data,  
a troubleshooting system can detect subtle anomalies in the program behavior of all \dlt workers. But, given the sheer volume of profiling data across the stack, analyzing the data of all workers at runtime cannot be done at scale, so it cannot serve as an online troubleshooting approach for production \dlts.

\para{Contributions.} \chang{We present \sys, the first online performance troubleshooting system that offers both fine-grained observability as in offline profiling and cluster-wide coverage as in online monitoring (\eg, can be enabled simultaneously at all \dlt workers to cover the entire GPU cluster), effectively localizing root causes of \dlt performance issues.}


\sys{} pinpoints root causes of performance issues in both hardware and software components.
It scales to modern AI models on large-scale GPU infrastructure---
identifying abnormal function executions for a 3,400-GPU \dlt only takes 3 minutes (\edit{with the ability to analyze a 1,000,000-GPU \dlt in 7 minutes}). \sys supports all training frameworks using PyTorch. To use \sys, one only needs to import it in the Python source code for \dlt without other modifications.

The key insight of \sys is to develop highly efficient differential observability of \dlt profiling data. 
\sys{} does not attempt to directly analyze raw profiling data of all workers (\eg, compare all function events between all workers).
Instead, \sys{} summarizes the {\it \pattern} of causally related \dlt functions on the critical path of the observed performance anomaly. 
We show that it is 
possible to concisely summarize each function's \patternss using statistical metrics (\eg, execution duration, hardware resource utilization, \etc), without loss of diagnosability. 
Since the size of \patternss is orders of magnitudes smaller than raw profiling data, the abnormal behavior can be efficiently identified by pattern comparison, leveraging the asymmetric architecture of model training.

We have developed \sys as a production service in our \dlt infrastructure with $\sim$100,000 GPUs and deployed it for 1.5 years. \sys has been helping our customers to troubleshoot various \dlt performance issues which can hardly be addressed by existing practices or state-of-the-art techniques, including hardware and code-level issues.

\para{Summary.} This paper makes the following contributions:
\begin{packed_itemize}
    \item We present \sys as the first profiling-based online performance troubleshooting system for \dlt in production.
    \item We describe the technique of summarizing \patternss of \dlt functions, which enables highly efficient differential observability for \dlt.
    \item We share our production results of \sys{}.
\end{packed_itemize}
\vspace{-12.5pt}
\section{Background and Motivation}
\label{sec-motiv}

\vspace{-7.5pt}
\subsection{Performance Issues in \dlt}
\label{subsec-motiv-current}

\begin{figure}[!t]
	\centering
	\includegraphics[width=\linewidth]{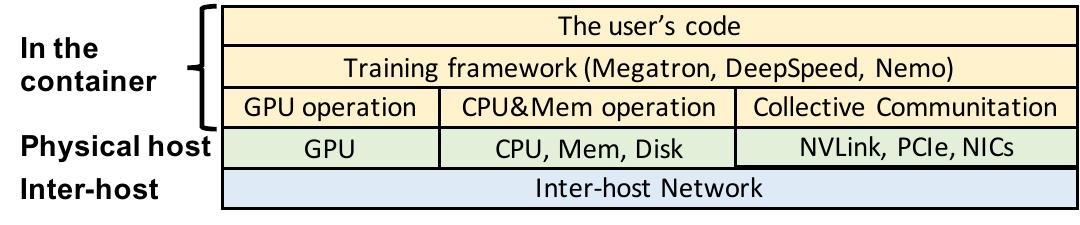}
 \vspace{-18pt}
	\caption{The full stack of large model training.}
 \vspace{-5pt}
	\label{fig:lmtexe}
\end{figure}

Large model training (\dlt) requires the cooperation of multiple hardware and software modules (Figure~\ref{fig:lmtexe}). The hardware typically includes GPUs~\cite{NVIDIA_H100, NVIDIA_H800}, CPUs, DRAM, PCIe, NVLinks~\cite{NVLINK}, NICs~\cite{CX6, CX7}, and inter-host networks~\cite{greenberg2009vl2,HPN}. Software modules include GPU operators, communication libraries like NCCL~\cite{NCCL}, training frameworks (\eg, Megatron~\cite{megatron}, Nemo~\cite{nemo}), containers, \etc. 
Malfunctions in any of the above components can lead to performance issues, such as throughput decrease, fluctuations, or training blockage. 
In this paper we do not focus on \dlt crashes, which can be explicitly identified by error messages.

\begin{figure}[!t]
	\centering
 \vspace{-10pt}
 \includegraphics[width=\linewidth]{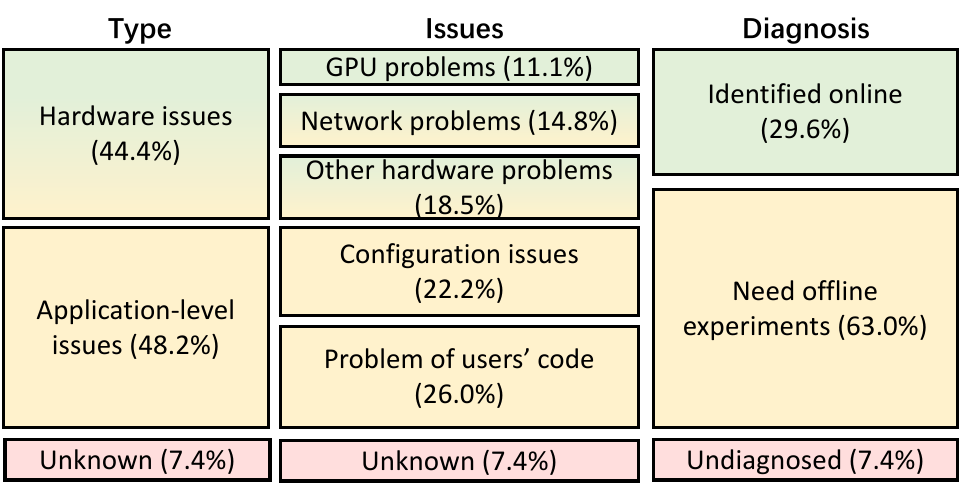}
   \vspace{-17pt}
	\caption{An overview of \dlt performance issues.}
   \vspace{-5pt}
	\label{fig:performance_issues}
\end{figure}

\camera{Figure~\ref{fig:performance_issues} breaks down all \dlt performance issues in our production \cluster with $\sim$100,000 GPUs over nine months.}
These issues manifest as slow computation or communication, but with diverse root causes---44.4\% of them are caused by hardware issues, and 48.2\% are software issues. 

Computation slowdowns can be caused by hardware (\eg, GPU throttling),
deficient user code, misconfigurations, or incompatible libraries. 
For example, in our customers' models, some training code frequently transfers data between CPUs and GPUs, introduces excessive synchronization, uses outdated PyTorch versions, poorly manages memory, or contains custom functions that run unexpectedly long.

Communication slowdowns can stem from physical networks, including NVLink, PCIe, NICs, and switches with optical modules. Although major communication throughput reduction is easy to detect, pinpointing the exact machines or hardware components is much more challenging. The difficulty is because most communications in \dlts are collective operations---a slowdown in any network link can proportionally reduce the communication throughput of many workers simultaneously. Besides, bugs in communication libraries, RDMA, and workload managers (\eg, Slurm~\cite{slurm}) can also lead to communication slowdowns.

Furthermore, system services used by \dlt can lead to performance issues as well. For example, storage services 
can affect the performance of data loading or checkpointing during \dlt. 
Management services like load tests and monitoring are co-located on each host; unexpected behavior can lead to resource contention, slowing down the \dlt. 


\vspace{-7.5pt}
\subsection{Common Practices}
\label{sec:practice}

\begin{table}
\centering
\vspace{-10pt}
\caption{Representative performance diagnosis tools for \dlt}
\label{tab:montitor_tools_overview}
\footnotesize 
\setlength{\tabcolsep}{3pt} 
\begin{tabular}{@{}ccccccc@{}}
\toprule
\multirow{3}{*}{\makecell[c]{Usage\ }} & \multirow{3}{*}{\makecell[l]{Tool}} & \multicolumn{4}{c}{Diagnostic Information} \\ \cmidrule(l){3-6}
 & & \makecell{GPU/DRAM/\\PCIe/NVLink} & NIC & \makecell{Python\\Events} & Kernels \\
\midrule
\multirow{5}{*}{\makecell[c]{Online}}
& DCGM~\cite{dcgm} & 1Hz & \XSolidBrush & \XSolidBrush & \XSolidBrush \\
& MegaScale~\cite{jiang2024megascale} & \XSolidBrush & 1kHz & \XSolidBrush & \Checkmark \\
& Dynolog*~\cite{dynolog} & 0.1Hz & 0.1kHz & \XSolidBrush & \XSolidBrush \\
& NCCL Profiler~\cite{ncclprofiler} & \XSolidBrush & \XSolidBrush & \XSolidBrush & \Checkmark \\
& eBPF~\cite{ebpf} & \XSolidBrush & \XSolidBrush & \XSolidBrush & \Checkmark \\
\midrule
\multirow{2}{*}{\makecell[c]{Offline}}
& Nsight Systems~\cite{nsys} & 10-200kHz & 1kHz & \XSolidBrush & \Checkmark \\
& Torch Profiler~\cite{torch_profiler} & \XSolidBrush & \XSolidBrush & \Checkmark & \Checkmark \\
\midrule
\multirow{1}{*}{\makecell[c]{Online}}
& \textbf{\sys} & 10-200kHz & 1kHz & \Checkmark & \Checkmark \\
\bottomrule
\end{tabular}
\vspace{-5pt}
\begin{flushleft}  
\footnotesize *Dynolog supports Torch Profiler as a plugin, to collect Python and kernel information. But, its diagnosis is based only on hardware information and thus is not regarded as a tool based on Python and kernel events.
\end{flushleft}
\end{table}

Before \sys, we relied on existing tools or techniques to troubleshoot performance issues of \dlt (Table~\ref{tab:montitor_tools_overview}).
However, none of them meets our needs in production.

\para{Monitoring.} Our GPU clusters widely deployed online monitors on hardware performance counters of GPUs, PCIe, NVLink, and the inter-host network with second-level sample rates, using existing techniques (\eg, in DCGM~\cite{dcgm}, Dynolog~\cite{dynolog}, and MegaScale~\cite{jiang2024megascale}).
\camera{There are also tools that collect function-level information online.
NCCL Profiler~\cite{ncclprofiler} (and related work such as Mycroft~\cite{deng2025mycroft} and Aegis~\cite{dong2025evolution}) focuses on collecting collective communication function events, whereas eBPF~\cite{ebpf} techniques (\eg, bpftrace~\cite{bpftrace}) collect events from system calls or other user-specified functions.}
However, as shown in Figure~\ref{fig:performance_issues}, \chang{only 29.6\% of performance issues can be diagnosed and root-caused by these tools.}

The key reason is that online monitors are coarse-grained and unaware of detailed behavior, creating 
a fundamental gap between observations and root causes. First, performance issues caused by configurations or code are invisible to hardware monitoring.
Second, most warnings from monitors \edit{are false positives~\cite{dong2025evolution}}, they do not necessarily indicate performance issues in \dlt; they can also be results of temporarily high pressure on hardware (\eg, excessive CNPs) or correctable errors (\eg, XID 63,64, and 92). 
Third, some hardware misbehaviors are fine-grained and bursty (\eg, <1ms), which hardware monitors at second-level granularity may miss. Finally, even simple issues, which are supposed to be detected by hardware monitors like link down, still have a chance to evade,
because in production, hosts are dynamically added or removed, making it challenging to achieve 100\% hardware monitoring coverage~\cite{dong2025evolution}.

In practice, engineers usually analyze the code behavior of \dlt in conjunction with hardware throughput and warning information, which have not been automated online.

\para{Profiling.} Profilers provide fine-grained observation of \dlt and are used as a supplement to online monitors. Nsight System and Torch Profiler record all activities in \dlt execution with microsecond-level hardware sampling, which provide sufficient information to identify the root cause of sophisticated performance issues. 

However, given their significant overhead (\eg, Torch Profiler generates 100+ MB profiling data
for one worker per second),
profiling can only be done in a small testbed, or for a few specific workers (\eg, only for rank~0) among thousands of workers in an online \dlt.
Engineers must first reproduce the observed performance issues in production in the testbed before using a profiler.
Unfortunately, reproducing production issues is not always possible or timely.
Besides problems that only manifest at a large scale, a practical problem is that
user code is often confidential and cannot be disclosed to the \cluster providers.
Engineers often have to reproduce production issues using similar open-source models,
which further reduces the chances of reproduction.
Although we invested tremendous efforts to reproduce \dlt performance issues reported by our customers,
there are still 7.4\% of performance issues that cannot be reproduced. For the ones that we eventually reproduced, it takes days to months of effort.

\vspace{-5pt}
\subsection{Goals and Challenges}
\label{subsec-motiv-goal}
Our goal is to significantly improve both efficiency and effectiveness of \dlt performance troubleshooting in production at scale, with
(1) {\it fine-grained observability:} profiling all function execution events during \dlt, with fine-granularity hardware sampling, 
 (2) {\it efficiency and scalability:} as fine-grained observability leads to large volumes of profiling data, online troubleshooting must be efficient and scalable to large \clusters;
(3) {\it low overhead:} 
introducing no performance impact on the routine \dlt training, and minimizing the requirements for computation power, network bandwidth, and storage.
We address the following challenges:

\para{Challenge~1: Overhead of fine-grained observability.}

\squishlist

\item {\it Profiling generates a huge amount of data.}
The size of profiling data of a single worker, including application-level tracing (\eg, the timeline of function execution) and hardware metrics (\eg, using nsys~\cite{nsys} for GPU, NVLink, and PCIe, in 10k~Hz), is typically 100~MB per second. Suppose an \dlt requires 10,000 GPUs, a profiler can generate $\sim$1~TB data per second, which is untenable. 

\item {\it Profiling may reduce \dlt performance.}
Profiling may reduce \dlt training throughput (\S\ref{subsec-overhead}). Most customers cannot afford periodic profiling for their \dlt jobs.

\squishend

\para{Challenge~2: Needles in the haystack.}

\squishlist

\item {\it Extracting important information.}
Fine-grained profiling data far exceeds the volume that can be efficiently analyzed in real time (\eg, 1~TB/s for a 10,000-GPUs \dlt). 
In fact, it is even hard to aggregate all the data; so, algorithms that assume seeing all the data are hardly practical. Dropping or downsampling is not ideal, as it may reduce important details and thus impair the diagnosability.

\item {\it Avoid expensive coordination.}
In production, hosts are not always perfectly synchronized in clock time. 
Even with protocols like Network Time Protocol (NTP)~\cite{mills1991internet}, the error could be $\sim$10 milliseconds.
Algorithms that compare timestamps of events between workers would not work---a lot of function executions are on the millisecond or even microsecond scale.
A decentralized algorithm is needed.

\squishend
\vspace{-5pt}
\section{Main Idea}
\label{sec-method}
\vspace{-5pt}



The key idea of \sys is to develop highly efficient differential observability of \dlt functions, leveraging homogeneous architectures of \dlt.
We use the term ``{\it function}'' to refer to any procedure in \dlt, including Python functions, GPU/CPU kernel functions, memory operations, \etc.
Compared to traditional systems, \dlt function executions exhibit two new characteristics.
\squishlist
\item Runtime behaviors of function executions \camera{(\eg, the start/end time of execution, and the hardware usage during execution)} are expected to be similar across workers \camera{or,
when not identical (\eg in heterogeneous computation like expert/pipeline parallelism), to follow a relatively stable distribution,} as a model usually has high symmetry to be distributed to multiple workers (examples in Appendix~\ref{appendix-behaviors}).
\item \chang{In our production, once a training job encounters a performance issue, it always persists until remediation. This is because training iterations repeatedly execute the same functions, and neither hardware faults nor code bugs are self-healing, so repeated execution continues to exhibit the same performance degradation.}
\squishend
Therefore, most performance issues can be diagnosed by observing abnormal function runtime behaviors in comparison to all other function executions.
We elaborate on this idea using a representative example of an \dlt performance issue caused by ring communication functions.

\begin{figure}[!t]
	\centering
	\includegraphics[width=\linewidth]{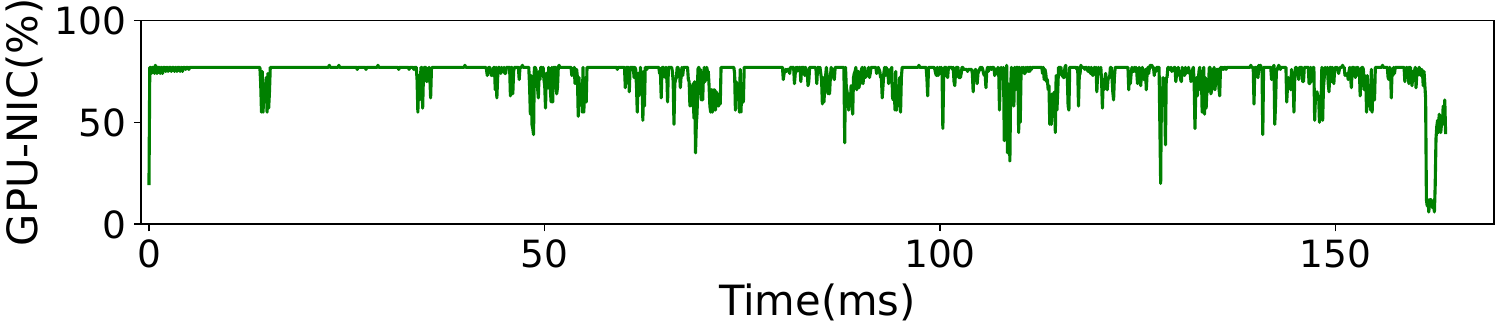}
 \vspace{-15pt}
	\caption{PCIe bandwidth utilization (from GPU to NIC) during a ring communication without network issues.}
	\label{fig:before throttling}
\end{figure}

 Ring communication in \dlt (like NCCL~\cite{NCCL} AllReduce and AllGather) connects workers in  ring-like topologies for direct data exchange with only their adjacent neighbors. A ring topology utilizes both NVLinks (intra-host) and GPU-NIC (inter-host) links.
In this example, a NCCL AllReduce group with 32 GPUs is distributed on 4 hosts, each hosting 8 GPUs, with every pair of GPUs sharing two bonded NICs. A malfunctioning NIC downgrades a bond by 50\% and creates a performance issue for the NCCL AllReduce function.
Since multiple rings typically share the intra-host network (\eg, NVLink) but use different NICs for inter-host communication, we focus on each worker's GPU-NIC throughput during NCCL AllReduce (Figure~\ref{fig:before throttling}), to distinguish the communication behaviors of different rings and different workers.

\begin{figure}[!t]
	\centering
	\includegraphics[width=0.6\linewidth]{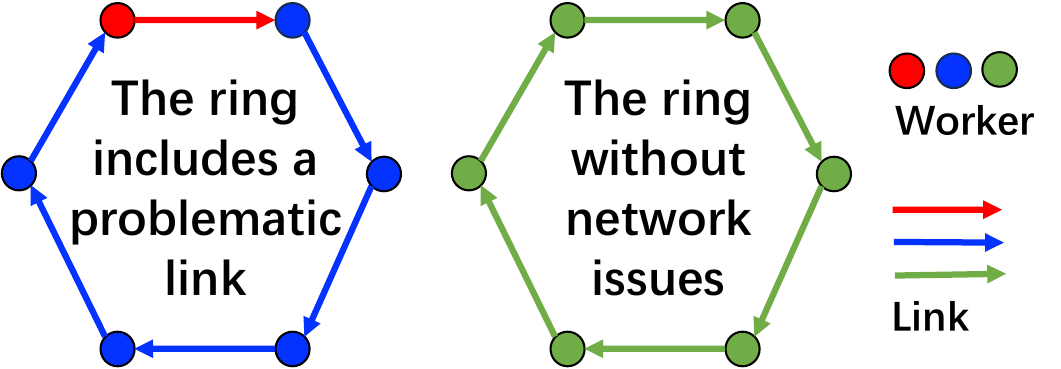}
	\caption{The ring topology with and without a problematic link. We illustrate a link with only six workers for simplicity. There are 32 in our experiment.}
	\label{fig:ring}
\end{figure}

\begin{figure}[!t]
	\centering
	\begin{subfigure}[b]{\linewidth}
		\centering
		\includegraphics[width=\linewidth]{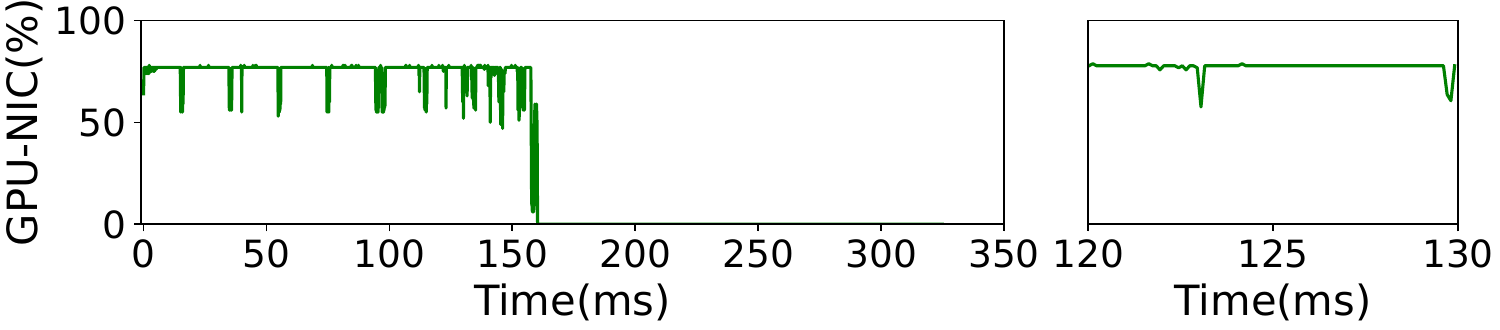}
		\caption{GPU-NIC links at maximal throughput.}
        \vspace{-5pt}
		\label{fig:good channel}
	\end{subfigure}
	
	\begin{subfigure}[b]{\linewidth}
		\centering
		\includegraphics[width=\linewidth]{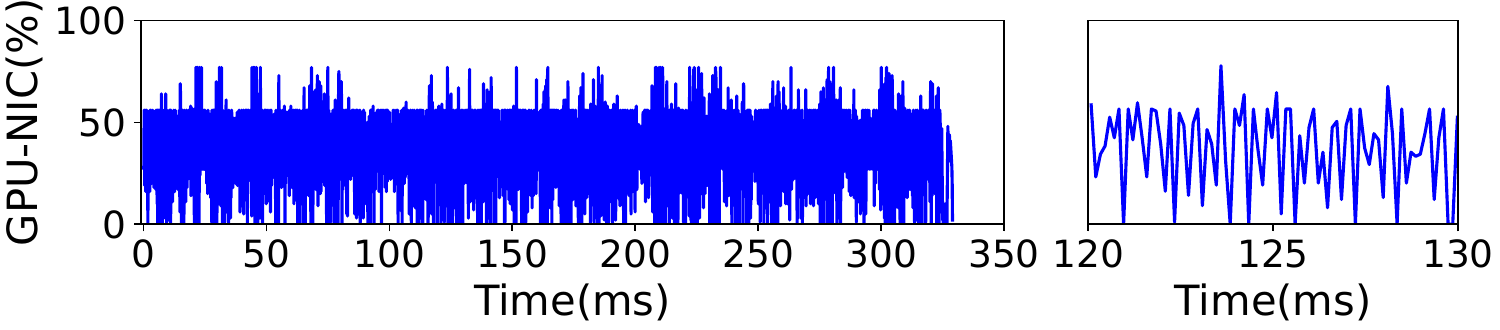}
		\caption{GPU-NIC links of lower throughput and high fluctuation.}
        \vspace{-5pt}
		\label{fig:affected channel}
	\end{subfigure}
	
	\begin{subfigure}[b]{\linewidth}
		\centering
		\includegraphics[width=\linewidth]{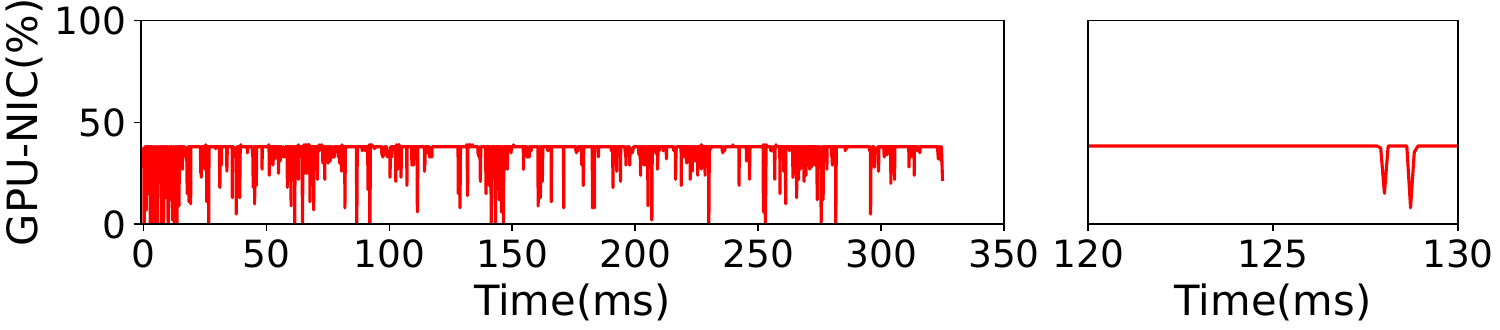}
		\caption{GPU-NIC link of lower and stable throughput (slow link).}
		\label{fig:root faulty channel}
	\end{subfigure}
	 \vspace{-20pt}
	\caption{GPU-NIC throughput pattern. Left figures are the bandwidth utilizations during the execution of a ring communication function, and right figures show enlarged views of local areas for better observation.}
 \vspace{-5pt}
	\label{fig:after throttling}
\end{figure}

After downgrading a network bond, all workers' GPU-NIC throughputs are categorized into three different patterns.
\squishlist
 \item For workers not included in a ring using this downgraded bond (green nodes in Figure~\ref{fig:ring}), the throughput pattern (Figure~\ref{fig:after throttling}\subref{fig:good channel}) is identical to the healthy state (Figure~\ref{fig:before throttling}). 
 \item For workers in a ring using the downgraded bond but not directly connecting to this bond (blue nodes in Fig.~\ref{fig:ring}), the average throughput is lower (Figure~\ref{fig:after throttling}\subref{fig:affected channel}) with fluctuations. 
 \item The worker that directly connects to the downgraded bond (red node in Figure~\ref{fig:ring}) also has low average throughput (Figure~\ref{fig:after throttling}\subref{fig:root faulty channel}) but is more stable than Figure~\ref{fig:after throttling}\subref{fig:affected channel}.
\squishend

\begin{figure*}[!t]
    \centering
    \includegraphics[width=\textwidth]{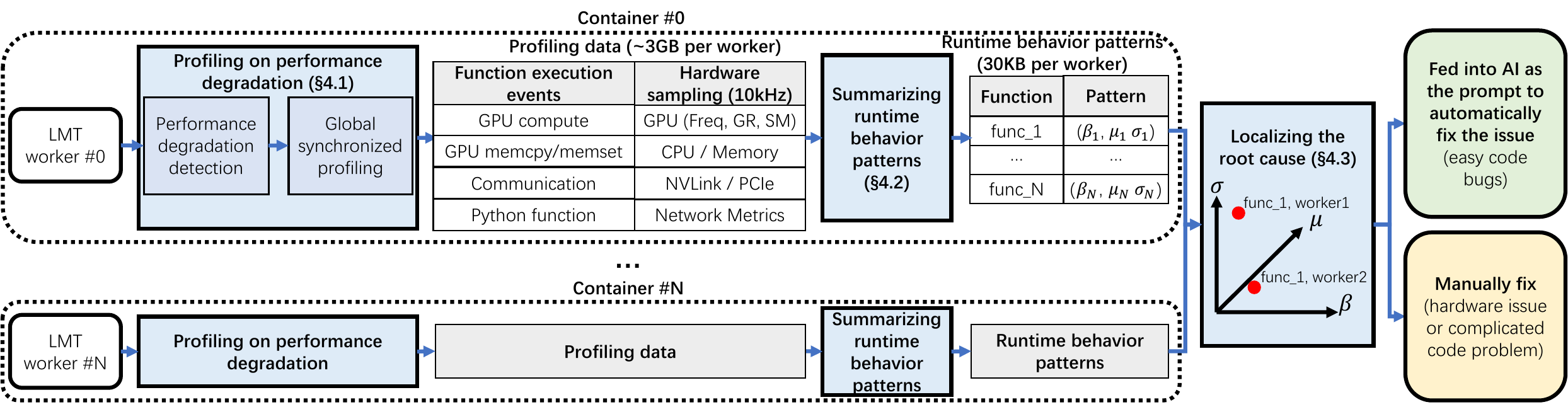}
    \caption{Overview of the \sys system.}
    \label{fig:overview}
    \vspace{-5pt}
\end{figure*}

To explain the low average throughput of Figures~\ref{fig:after throttling}\subref{fig:affected channel} and \ref{fig:after throttling}\subref{fig:root faulty channel} compared with Figure~\ref{fig:after throttling}\subref{fig:good channel}: 
during the ring communication, the NCCL communication library constructs multiple rings, each using different NICs, to link all workers head-to-tail. If a ring contains no slow link, all of its links operate at maximal throughput. For rings including a slow link, the average throughput of all workers included in this ring drops. 

To explain the fluctuation in Figure~\ref{fig:after throttling}\subref{fig:affected channel}:
ring communication always transfers data in small chunks one by one. At each stage, each worker transfers a chunk to the next worker, and does some computation based on the chunk it received. With a slow link in a ring, although other links can still reach maximal throughput, they have to wait for the slow link after each stage of chunk transmission. So their throughput fluctuates between zero and the maximal, with a high standard deviation. The slow link does not need to wait, so its throughput remains low with a small standard deviation \camera{as shown in Figure~\ref{fig:after throttling}\subref{fig:root faulty channel}}. 

\para{Insights.} The above example demonstrates three important opportunities: (1) performance issues can be observed by profiling the behavior of function executions\edit{, and we can use differential observability to localize} the offending function executions with abnormal behavior (\eg, low average GPU-NIC throughput without fluctuation);
and (2) we do not need to analyze fine-grained raw observability data of all the functions; instead, we only need to summarize their {\it \patternss}.  
In the ring communication example, each worker only needs to provide two numbers (mean and standard deviation of GPU-NIC throughput) as the pattern of the ring communication function;
\edit{(3) Fortunately, {\it \patternss} (\eg, mean and standard deviation) can be independent from absolute timestamps, which can be compared across multiple hosts without any clock time synchronization, solving the Challenge~2 in \S\ref{subsec-motiv-goal}.} 
In summary:

\begin{tcolorbox}[boxsep=1mm,left=1mm,right=1mm,top=1mm,bottom=1mm]
Most (if not all) \dlt performance issues can be diagnosed by highly efficient differential observability on \patternss of individual \dlt functions.
\end{tcolorbox}

\section{\sys Design}
\label{sec-design}
\vspace{-5pt}

\sys is an online performance troubleshooting system for large-scale model training (\dlt).
It embodies the insights discussed in \S\ref{sec-method} in a production service for customers who run \dlt on our \clusters.
\sys employs a highly efficient approach to detect performance issues by online profiling on demand, introducing little overhead on \dlts.
When a performance issue is detected, \sys automatically troubleshoots it.
\sys uses online profiling to collect fine-grained information, but it does not aggregate raw profile data of every function; instead, \sys summarizes the \patternss of each function at every worker node and aggregates only the concise behavior patterns.
The diagnosis is performed on the aggregated behavior patterns based on the differential observability principle.

Figure~\ref{fig:overview} gives an overview of \sys.
\sys has three main components: 
(1) detecting performance degradation of \dlt to trigger online profiling simultaneously for all workers in \dlt,
(2) summarizing \patternss of each function from raw profiling data,
and (3) a centralized localization algorithm that pinpoints the root-cause function based on the behavior patterns.
Both (1) and (2) are deployed per worker,
while (3) is global.

\begin{figure}[!t]
	\centering
	\includegraphics[width=\linewidth]{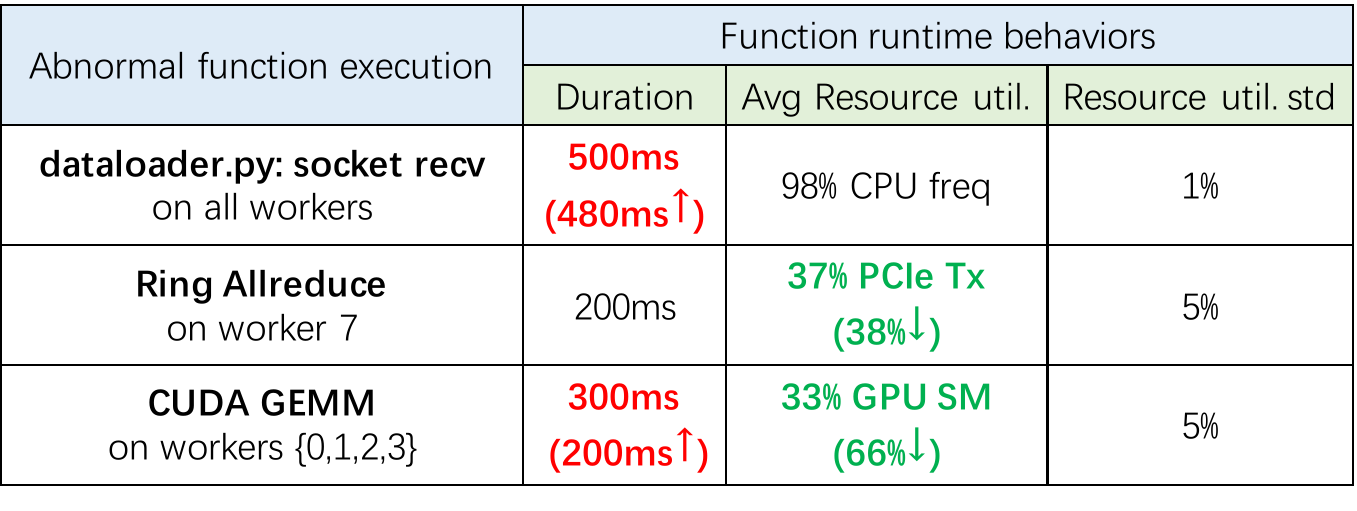}
	\vspace{-15pt}
        \caption{An example of \sys output.}
        \vspace{-10pt}
	\label{fig:output}
\end{figure}

\para{Usage.} To deploy \sys, \cluster
providers do not need access to the user code. The user only needs to import \sys in the \dlt code (\texttt{import \sys});
no other code change is needed. 
Behind the line of import, \sys initializes the detector of performance degradation, registers handler functions to start profiling, and creates a daemon process for summarizing and uploading behavior patterns.
Figure~\ref{fig:output} illustrates the output of \sys. As a function-centric approach,
\sys pinpoints which functions on which workers are executed abnormally, and describes how they behave differently from expectation or from other workers. 
In Figure~\ref{fig:output}, all workers encounter slow socket receive in data loading, indicating slow storage I/O. Worker~7 is slow in collective communication, indicating a network degradation on its connected link. In addition, worker 0,1,2, and 3 execute \texttt{\small GEMM} (a CUDA kernel function~\cite{gemm}) slowly because of not fully utilizing GPUs, indicating the problem of GPU throttling.

\sys's output is directly fed into an AI assistant to try automated fixing: for simple code bugs, the assistant may automatically patch the code (a real case in \S\ref{subsec-case5}), whereas hardware faults or complex code issues typically still require human effort to fix (\eg, replacing the problematic hardware).

\vspace{-5pt}
\subsection{Profiling on Performance Degradation}
\label{subsec-design-collector}
\vspace{-2.5pt}

Profiling an \dlt task online introduces overhead. 
To minimize the overhead while collecting profiling data timely when performance issues occur, \sys automatically detects performance degradation \camera{and then performs profiling simultaneously for all workers, ensuring the problematic behavior is} \chang{included in} \camera{the profiling data.}

For an \dlt task, the time spent on one training iteration is the key performance metric. \sys devises a solution to monitor the training time of each iteration without accessing user code, and then triggers online profiling if the iteration time is abnormal. Note that approaches that require accessing user data (such as logs) do not work with our usage model.

\begin{figure}[!t]
	\centering
	\includegraphics[width=1.02\linewidth]{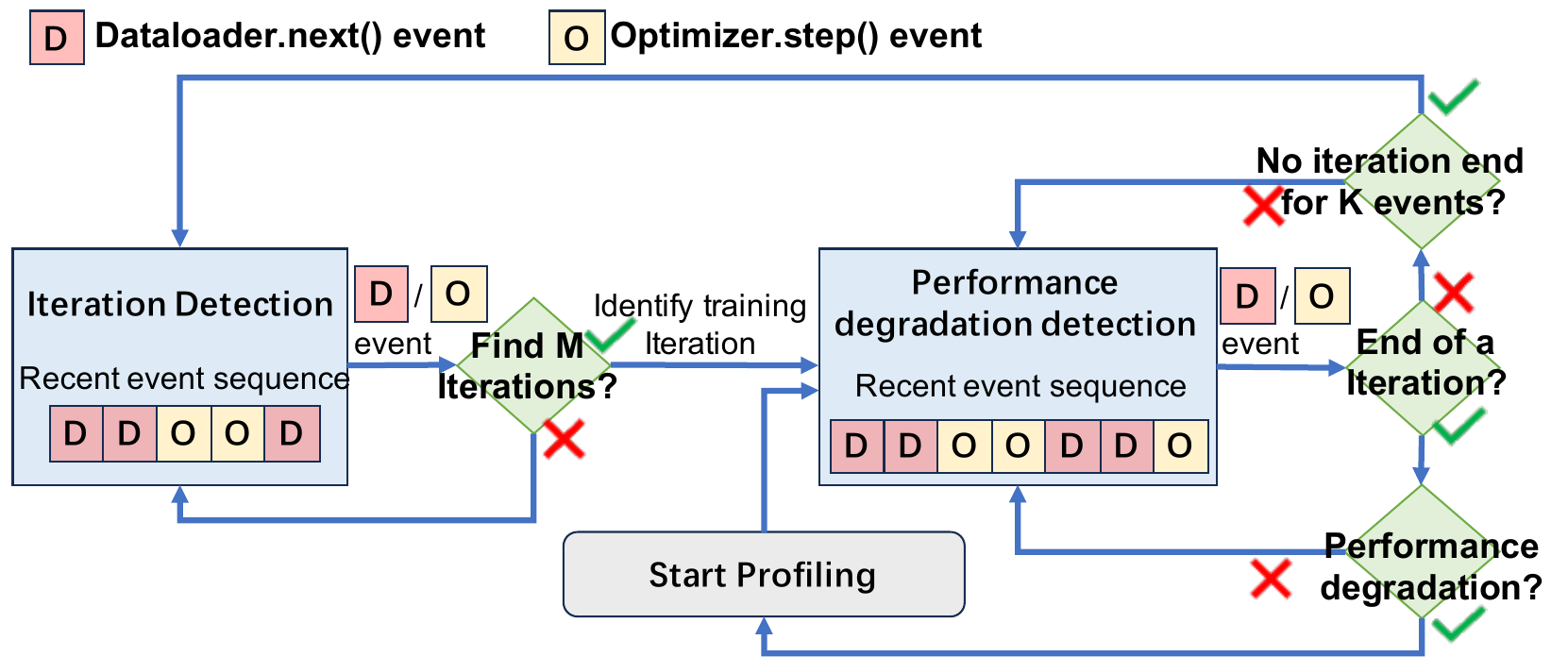}
  \vspace{-10pt}
	\caption{\sys's performance degradation detection.}
 \vspace{-5pt}
	\label{fig:automation}
\end{figure}

\para{Indicators of iteration time.}
A PyTorch training iteration always involves several \texttt{\small dataloader.next()} calls, followed by several \texttt{\small optimizer.step()} calls (the number depends on training parameters). 
In this sequence of events, the duration from the first \texttt{\small dataloader.next()} to the last \texttt{\small optimizer.step()} is regarded as the duration of a complete training iteration. 

When it is \texttt{\small import}-ed, \sys wraps the two functions by adding time counters. 
Basically, \sys replaces the two PyTorch functions with the wrapped versions. Since the two functions are Python functions (which are not compiled before \dlt executions), the replacement is done at runtime.

\para{Iteration detection.}
During an \dlt, \sys continuously records event sequences of \texttt{\small dataloader.next()} and \texttt{\small optimizer.step()}. After detecting $M$ (=10 in practice) identical sequences starting with \texttt{\small dataloader.next()} and ending with \texttt{\small optimizer.step()}, this sequence is defined as the \emph{training iteration sequence}.

\para{Performance degradation detection.}
\sys continuously monitors incoming events of \texttt{\small dataloader.next()} and \texttt{\small optimizer.step()} with the identified training iteration sequence. Each time \sys successfully matches a complete training iteration sequence, it records the duration of that iteration. \sys considers that performance degradation has occurred in the following two situations: (1) The average duration of the recent $N$ (=50 in practice) iterations exceeds the recent shortest iteration time by more than 5\%. (2) The current training iteration sequence has not yet been fully matched, but the time elapsed since the last event received is at least $5\times$ the average iteration duration (indicating the training is blocked). If \sys fails to match a training iteration after $K$ (=200 in practice) consecutive event receptions, it goes back to the previous iteration detection phase to redetect the training iteration sequence. While this does not often happen in common \dlt programs, it significantly enhances the robustness of \sys when users implement special functionalities in their code, to make sure the algorithm always works. Figure~\ref{fig:automation} illustrates the procedure.

\para{Global synchronized profiling.}
Each \dlt worker connects to an \sys daemon. When \sys detects a slowdown, it notifies all daemons via TCP; each daemon signals its worker to invoke a pre-registered handler, ensuring profiling runs in the \dlt main thread (required by some APIs, \eg, CUPTI in Torch Profiler).

\camera{\sys ensures synchronized profiling across workers via iteration ID: rank-0 continuously report the current iteration ID; upon a profiling trigger, the rank-0 daemon computes unified start/stop iteration IDs for profiling (with the start set a few steps ahead to ensure no worker would miss it), and other daemons periodically poll these IDs and start/stop profiling accordingly.}

A profiling session lasts for 20 seconds by default (configurable). We use Torch Profiler to collect execution events (Python/CPU ops, memory ops, and CUDA kernels), and \texttt{nsys} to sample hardware metrics at 10\,kHz (GPU, DRAM, NVLink, PCIe, and network). Each daemon aggregates and collects the profiling data.

\subsection{Summarizing Behavior Patterns}
\label{subsec-design-extractor}

\begin{figure}[!t]
	\centering
	\includegraphics[width=\linewidth]{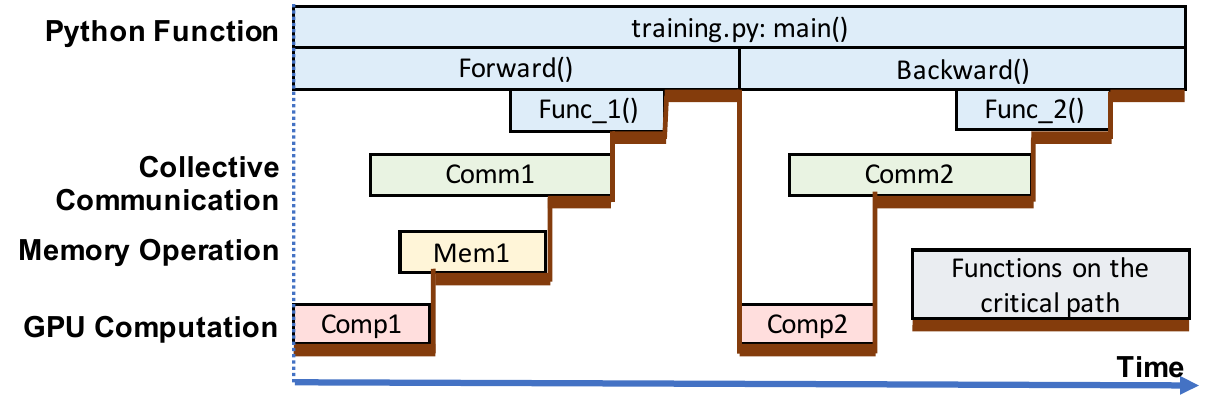}
 \vspace{-17pt}
	\caption{The critical path of \dlt performance.}
	\label{fig:critical_path}
 \vspace{-5pt}
\end{figure}

\sys summarizes behavior patterns of each function at the worker level.
Note that \sys only concerns functions on the critical path of \dlt (Figure~\ref{fig:critical_path})---only these functions contribute to the end-to-end performance, despite the huge number of function executions in an \dlt.

\sys's solution is (1) finding the function execution events on the critical path, including GPU computation kernels, collective communication functions, memory operations, Python functions, and all other functions executed in \dlt; and (2) clustering all execution events of each function (for Python functions, the entire call stack must be identical to be considered the same function), then defining several patterns to summarize the behavior of each function.

\para{Critical path.}
As shown in Figure~\ref{fig:critical_path}, \dlt\ assigns different types of function with priorities (higher is more critical):
GPU compute kernels $>$ memory operations (\eg, malloc, memcpy) $>$ collective communication kernels (\eg, NCCL AllReduce) $>$ Python functions.
A function’s execution (or a subinterval of it) is on the \dlt's critical path iff no higher-priority function is executing during that time.
For Python functions, we further require that they run in the training thread (\eg, not spawned by \texttt{\small \_bootstrap}) and have no executing child calls.

The rationale for such a definition is that a well-optimized \dlt should try to keep GPUs busy, so in general we should focus on GPU kernel executions and function executions during GPU idle time. 
Therefore, all function executions are prioritized based on their relevance to GPU computation. Only when a function unrelated to GPU computation fails to overlap with GPU computation (indicating it blocks GPU computation), it should be regarded as a possible problematic function. Reducing its execution time can improve end-to-end training performance. Otherwise, it remains irrelevant to performance since it does not block GPU computation.

\camera{Note that one could craft a counterexample against \sys, for example, by fully overlapping communication with GPU computation, making communication stalls invisible on the critical path. However, in production \dlt, computation and various communications have strict dependencies. While some communication can overlap with computation, we have never seen real workloads where all communication is fully overlapped; thus it still partially appears on the critical path and can be detected by \sys.}

\para{Runtime behavior patterns.}
For each function $f$ running on worker $w$, we define its pattern of behavior $P_{f, w}$ as a 3-dimensional vector:
\begin{alignat}{2}
P_{f, w}=(\beta_{f, w}, \mu_{f, w}, \sigma_{f, w})\label{equ:pattern}
\end{alignat}

$\beta_{f, w}$ represents the percentage of the profiling window that function $f$ spends time on the critical path of worker $w$.
\begin{alignat}{2}
\beta_{f, w} = \frac{\int_{t \in T}{C(f, w, t)}}{|T|}\label{equ:ratio}
\end{alignat}
where $T$ is the duration of the profiling window (\eg, 20 seconds in \S\ref{subsec-design-collector}), and 
\begin{alignat}{2}
C(f, w, t) = \begin{cases} 
1 & \text{$f$ is on $w$'s critical path at time $t$}, \\
0 & \text{Otherwise } .
\end{cases}
\end{alignat}
Base on the definition, we have $\beta_{f, w} \in [0, 1]$.

$\mu_{f, w}$ represents the average hardware resource utilization during all executions of function $f$ on worker $w$. In this context, ``resource'' refers to hardware resources that determine performance of function $f$. For example, GPU computation kernels correspond to GPU SM frequency, Python functions to CPU utilization, intra-worker collective communication to NVLink utilization, and inter-worker collective communication corresponds to the bandwidth utilization between GPU and NIC, due to the significantly lower NIC bandwidth compared to intra-worker NVLink bandwidth.
\begin{alignat}{2}
\mu_{f, w} = \frac{\sum_{e \in E_{f, w}}{|L(e)|avg_U(L(e))}}{\sum_{e \in E_f}{|L(e)|}}\label{equ:avg}
\end{alignat}
where $E_{f, w}$ represents all execution events of function $f$ on worker $w$ in the time window, so $e$ represents each one of them. $L(e)$ represents the {\it critical execution duration} of $e$. $avg_U(L(e)) \in [0,1]$ represents the average resource utilization during the critical execution duration. We have $\mu_{f, w} \in [0,1]$.

\begin{figure}[!t]
	\centering
	\includegraphics[width=\linewidth]{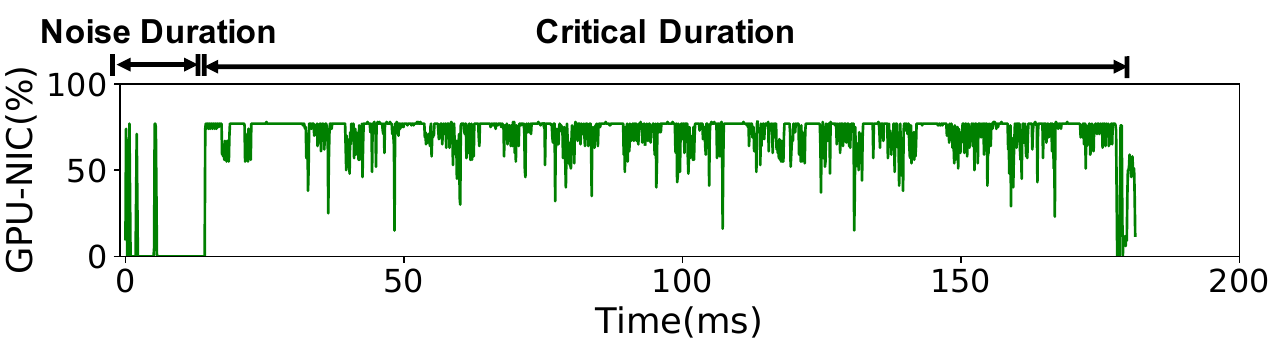}
  \vspace{-10pt}
	\caption{Critical duration of a collective communication function. A worker enters the communication early often need to wait for others, leaving several empty intervals of resource usage (noise). We only consider the critical duration in $\mu_{f, w}$.} 
	\label{fig:critical_duration}
\end{figure}

The definition of critical execution duration, $L(e) = [l_c, r_c]$, is not trivial, and is a subinterval of $f$'s function execution duration $[l, r]$, as shown in Figure~\ref{fig:critical_duration}. This is because in practical \dlt tasks, particularly within collective communication functions, numerous synchronization operations are involved. Workers that enter the collective communication kernel earlier usually perform a part of communication and then wait for others. Consequently, the throughput of network transmission are not continuous. The average resource utilization over the entire function execution does not accurately reflect the real communication performance. Thus, $L(e)$ is defined as the longest interval with intense resource usage among the whole function execution duration. We apply Algorithm~\ref{alg:max_gap_subinterval} to identify $L_e$ with discrete sampling of hardware resource usage.

\begin{algorithm}[t]
\footnotesize
\caption{Finding the critical execution duration}
\label{alg:max_gap_subinterval}
\begin{algorithmic}[1]
\REQUIRE Resource utilization samples $U = [U_1, U_2, \ldots, U_n]$ of function $f$'s execution duration $[l, r]$, where $U_i \in [0,1]$ for $\forall i$
\ENSURE Critical execution duration $[l_c, r_c]$

\STATE Total resource utilization $S = \sum_{i=1}^{n} U_i$
\STATE Initialize search bounds: $g_{\text{left}} \leftarrow 0$, $g_{\text{right}} \leftarrow n$
\WHILE{$g_{\text{left}} \leq g_{\text{right}}$}
    \STATE $g \leftarrow$ midpoint of $[g_{\text{left}}, g_{\text{right}}]$
    \STATE \textbf{if} there exists a subinterval $[l', r']$ satisfying:
    \begin{enumerate}
        \item $\sum_{i=l'}^{r'} U_i \geq 0.8 \times S$
        \item No more than $g$ consecutive zeros in $U_{l'}, \ldots, U_{r'}$
    \end{enumerate}
    \STATE \quad Update $g_{\max} \leftarrow g$, $[l_c, r_c] \leftarrow [l', r']$
    \STATE \quad $g_{\text{right}} \leftarrow g - 1$
    \STATE \textbf{else}
    \STATE \quad $g_{\text{left}} \leftarrow g + 1$
\ENDWHILE
\RETURN Subinterval $[l_c, r_c]$
\end{algorithmic}
\end{algorithm}

$\sigma_{f, w}$ represents the standard deviation of resource usage during all function executions of function $f$ on worker $w$, and its definition is similar to $\mu_{f, w}$:
\begin{alignat}{2}
\sigma_{f, w} = \frac{\sum_{e \in E_{f, w}}{|L(e)|std_U(L(e))}}{\sum_{e \in E_f}{|L(e)|}}\label{equ:avg}
\end{alignat}
and we also have $\sigma_{f, w} \in [0,1]$.

\edit{Note that all three dimensions of $P_{f, w}$ are independent of absolute timestamps. For example, to compute $\beta_{f, w}$, we only need the time difference between the start and end of function execution. This design is the key to performing inter-host pattern comparison without the need for clock synchronization.} 

\para{Data size of patterns.}
\begin{figure}[!t]
	\centering
	\begin{subfigure}[t]{0.49\linewidth}
		\centering
		\includegraphics[width=\linewidth]{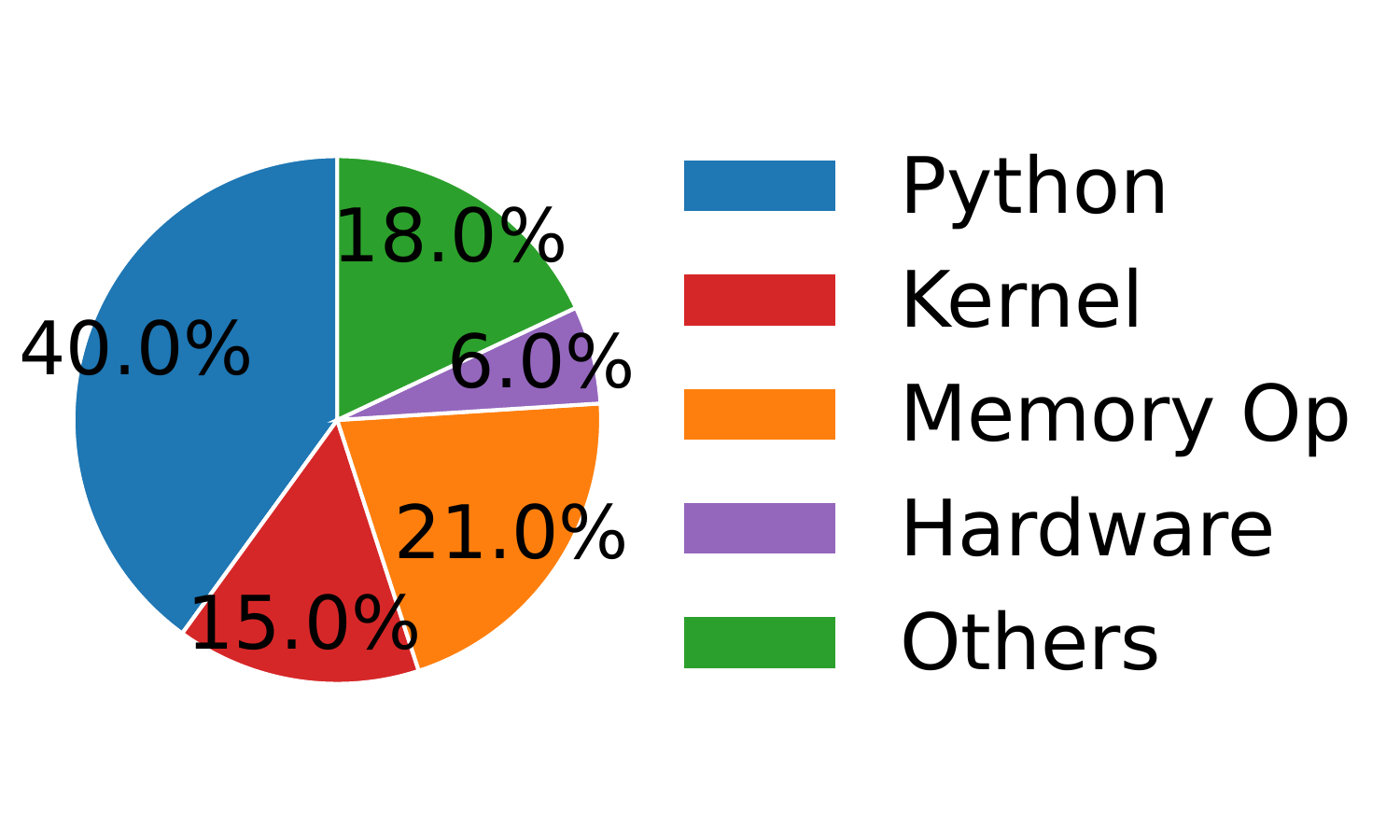}
		\caption{Raw profiling data (3GB)}
		\label{fig:raw_data}
	\end{subfigure}
	\hfill
	\begin{subfigure}[t]{0.49\linewidth}
		\centering
		\includegraphics[width=\linewidth]{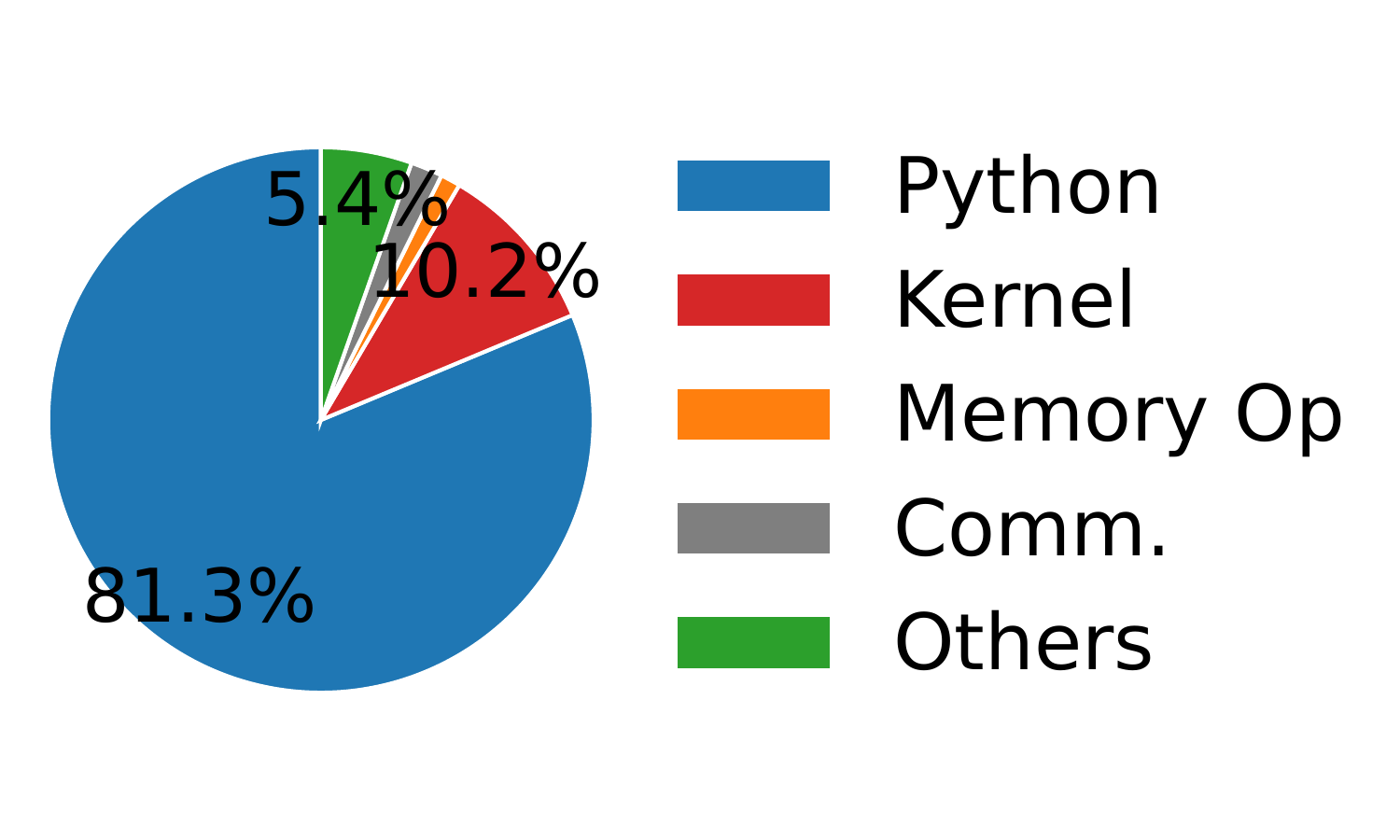}
		\caption{Runtime behavior patterns (30KB)}
		\label{fig:pattern}
	\end{subfigure}
  \vspace{-10pt}
	\caption{Data size of a single worker's \patternss compared with its raw profiling data.}
 \vspace{-5pt}
	\label{fig:pattern_size}
\end{figure}
Figure~\ref{fig:pattern_size}\subref{fig:pattern} shows the size of \patternss (30~KB) is $10^5 \times$ smaller than the raw profiling data ($\sim$3~GB). Patterns of Python functions contributes to the majority of the data volume, because \sys records the full call stack of each function, which is long (sometimes the call stack of a Python function includes 1,000 letters).

\vspace{-2.5pt}
\subsection{Localization}
\label{subsec-design-analyzer}

\sys localizes the abnormal function behavior based on
the \patternss in \S\ref{subsec-design-extractor}. \camera{In most cases, the abnormal function behavior can directly pinpoint a single plausible root cause. For more complex cases, our engineers make the final call, optionally with the help of an AI assistant.}

In \dlt execution, all performance issues can be categorized into two types: (1) a common problem of all \dlt workers, like hardware misconfigurations and low-efficiency code implementation, or (2) a special problem on only a part of workers, like hardware issues.
Correspondingly, given function $f$ executed on worker $w$, \sys defines two types of distances, (1) \emph{Distance from expectation} and (2) \emph{Differential distance}, to measure its abnormality. 

\para{Distance from expectation.} The first is the distance from expectation, $D_{f, w}$. In a well-optimized \dlt, we have an expected range for the \patternss of most functions (\eg, a Python function $f$ are expected to have a close-to-zero value of $\beta_f$ because \dlt should not be bottlenecked by CPU). 
Any function with a \pattern far from this range can be considered abnormal.
When there are a lot of workers with large distance from expectation, it indicates common issues in the \dlt, such as low hardware configurations across the cluster or deficient implementations of certain functions deployed on all workers.

To define the distance from expectation, we first need an expected range of patterns for function $f$, $R_f$:
\begin{alignat}{2}
R_{f} =  [\beta_f^l, \beta_f^r] \times [\mu_f^l, \mu_f^r] \times [\sigma_f^l, \sigma_f^r] \label{equ:range}
\end{alignat}
$R_f$ is assigned based on our production experience. For example, if $f$ is a Python function, we set $R_{f} =  [0, 0.01] \times [0, 1] \times [0, 1]$, \camera{since customers typically regard performance fluctuations within 1\% as measurement noise; deviations beyond this threshold are viewed as substantial regressions and thus trigger diagnostic requests.} If $f$ is a collective communication function, we set $R_{f} =  [0, 0.3] \times [0, 1] \times [0, 1]$. For GPU computation kernel functions, we set $R_{f} =  [0, 1] \times [0, 1] \times [0, 1]$, \ie, \camera{we expect a \dlt to spend most of its time on GPU computation to fully utilize the GPUs.}

Then we can define $D_{f, w}$ as the minimal Manhattan distance from $P_{f, w}$ to $R_f$:
\begin{alignat}{2}
D_{f, w} = \min_{p \in R_f}{d_{\text{Manhattan}}{(P_{f, w}, p)}}\label{equ:de}
\end{alignat}

\para{Differential distance.} The second is the distance to other workers, $\Delta_{f, w}$. In a large-scale \dlt, the behaviors of different workers should be highly consistent. If some workers exhibit significantly different function behaviors compared to the same functions executed on other workers, it often suggests that these workers are performing special operations that slow down overall training throughput (\eg, different in $\beta$-dimension), or that the hardware used by these workers for the function has performance issues that need to be addressed (\eg, different in $\mu$ or $\sigma$-dmension).

To make comparison between workers, first we make a max normalization to $P_{f, w}$:
\begin{alignat}{2}
\hat{P_{f, w}} = (\frac{\beta_{f, w}}{\max_{w \in W}{\beta_{f, w}}}, \frac{\mu_{f, w}}{\max_{w \in W}{\mu_{f, w}}}, \frac{\sigma_{f, w}}{\max_{w \in W}{\sigma_{f, w}}})
\end{alignat}

Then we define $\Delta_{f, w}$ to indicate how many workers have different patterns of $f$'s execution from worker $w$:
\begin{alignat}{2}
\Delta_{f, w} = \frac{\sum_{w' \in W_N}{I(\hat{P_{f, w}}, \hat{P_{f, w'}})}}{N} \label{equ:dp}
\end{alignat}
where $W_N$ is a subset of $N$ workers randomly sampled from the set of all workers $W$. In practical we set $N = \min{(100, |W|)}$. $I$ is a function indicating whether the distance between two vectors are below $\delta$:
\begin{alignat}{2}
I(x, y) = 
\begin{cases} 
0 & \text{if } d_{\text{Manhattan}}(x, y) < \delta, \\
1 & \text{otherwise.}
\end{cases}
\end{alignat}
In practical, we set $\delta$=0.4 based on our experience.

Note that our definition of $\Delta_{f, w}$ is not based on how far a worker is from other workers (\eg, simply average the distance to other workers), but how unique a worker's behavior is. This is because the pattern $P_{f, w}$ is in three dimensions, and each dimension has different physical meaning. A larger Manhattan distance does not mean a worker's behavior is more abnormal (\S\ref{sec-method} shows a classical example). 


\para{Localizing abnormal function executions.}
With the two distances, a function $f$ running on worker $w$ is abnormal 
if:
\begin{alignat}{2}
\beta_{f, w}>0.01\ \land \ (D_{f, w}>0 \ \lor \ \Delta_{f, w}> M_f + k MAD_f)
\label{equ:abnormal}
\end{alignat}
where $M_f = median\{\Delta_{f, w'} | w' \in W\}$ is the median value of distance to peers, and $MAD_f = median\{ |\Delta_{f, w'} - M_f|\ |w' \in W\}$ is Median Absolute Deviation~\cite{mad}, a robust measure of statistical dispersion. $k$ is set to 5.

$\beta_{f, w}$>0.01 means the function execution contributes at least 1\% to end-to-end performance. 
For most \dlt tasks, no more than 20 functions satisfy this requirement. $D_{f, w}>0$ indicates the runtime behavior of function $f$ is unexpected based on our production experience, and $\Delta_{f, w}> M_f + k MAD_f$ indicates the value of $\Delta_{f, w}$ is significantly larger than most other workers.

As shown in \S\ref{subsec-design-extractor}, the pattern of a worker is only $\sim$30~KB, even for an \dlt with 10,000 workers, the localization algorithm only consumes 300~MB data to output the final result, which is highly efficient to be executed on a single CPU core.

\para{Alternatives.} 
To localize abnormal function executions, we also tried existing clustering algorithms, including DBSCAN~\cite{ester1996density}, HDBSCAN~\cite{mcinnes2017hdbscan}, Gaussian Mixture Model~\cite{bishop2006pattern} and Mean shift~\cite{comaniciu2002mean}. They do not work well in production because they have at least one of the following limitations: (1) failing to distinguish noises and outliers, which is important in our scenario, or (2) having too many hyper-parameters, reducing their general applicability in practical cases.
\vspace{-5pt}
\section{Implementation}
\label{sec-impl}
\vspace{-2.5pt}

We built \sys with 7K lines of Python code, and deployed \sys 
as a service for our production \clusters. We
address the following practical challenges.


\para{Optimizations of profiling data generation.}
\sys relies on Torch profiler to generate a part of raw profiling data (\S\ref{sec-design}). 
However, Torch profiler introduces overhead to \dlt: (1) When the profiling window ends, it requires a long time to prepare and dump the data, and (2) after profiling, it leaves a lot of resources (\eg, hooks for CUDA functions added by CUPTI~\cite{cupti}) in the \dlt. They may reduce the performance of CUDA kernel execution even after profiling is finished.

\sys addresses the problems by optimizing the Torch profiler. First, the Torch profiler always transfers profiling data to Chrome tracing format, and then dumps the data using Kineto APIs~\cite{kineto}. Since Kineto supports dumping data with the same format, we remove the redundant and slow format transformation performed by Torch profiler, and directly apply Kineto to dump data. This reduces the time spent on data generation by 33\%. Second, \sys calls \texttt{\small cuptiFinalize()} to clear all remaining resources in \dlt after profiling. 

\para{Accessing hardware information within the user's container.}
In production \clusters, an \dlt typically runs inside containers; so does \sys daemons (\S\ref{subsec-design-collector}). However, \cluster providers usually restrict the permissions of user containers---\sys cannot directly access hardware information. \sys utilizes Kubernetes~\cite{kubernetes}'s
native ``emptyDir'' feature to share directories between containers, performing high-frequency hardware sampling through a privileged management container to place data in the shared directory. Therefore, \sys can obtain hardware profiling data within user containers without loosening permissions of user containers.

\para{Cooperating with other hardware monitors.}
In most production \clusters, coarse-grained hardware monitors are deployed on each physical host, using APIs such as DCGM and PCM to collect hardware metrics in real-time for health checks. However, some information (\eg, GPU metrics) can only be subscribed by one process at a time.
\sys's data generator cooperates with the host's monitoring system via signals (through a shared directory).
As \sys's profiling typically lasts about 20 seconds each time, we do not encounter any conflicts caused by the cooperation.



\vspace{-5pt}
\section{Evaluation in Production}
\label{sec-eval}
\vspace{-5pt}

\sys has been deployed in our production \clusters with \camera{$\sim$100,000} GPUs for 1.5 years. During this period, there were in total 80 serious \dlt performance issues for which we failed to localize the root causes using state-of-the-art techniques described in \S\ref{sec:practice} (see
Table~\ref{tab:issues}). 
\sys successfully identified root causes of 78 of them (success ratio=97.5\%), improving the training throughput of these \dlts (the largest one includes 6,144 GPUs) by $\sim$20\% to $\sim$100\%.

We present how we used \sys to diagnose three \dlts with 3,072, 3,400, and 128 GPUs, respectively, to show:

\squishlist
\item \sys can identify real-world performance issues of \edit{mixture of multiple hardware and code problems}. (\S\ref{subsec-case1}-\ref{subsec-case2})
\item \sys's output can be directly fed into AI as the prompt to help customers automatically fix the code bug. (\S\ref{subsec-case5})
\item \sys makes diagnosis in high efficiency, and does not introduce overhead to the routine training of \dlt. (\S\ref{subsec-overhead})
\squishend

In addition, we present two more case studies in Appendices \ref{subsec-case3} and \ref{subsec-case4}, including one of the only two (out of 80) cases where \sys failed to diagnose the issue. In Appendix \ref{state-of-the-arts}, we systematically compare \sys with state-of-the-arts.

\begin{table}[!t]
\vspace{-7.5pt}
\caption{80 serious performance issues identified by \sys{} (ONLY includes those not identified by our existing systems)}
\vspace{-2.5pt}
\centering
\small
\begin{tabular}{@{}llc@{}}
\toprule
Category & Root cause & Number of \dlt cases\\
\midrule
\multirow{3}{*}{Hardware issues} & GPU & 2 \\
                                 & CPU & 2 \\
                                 & Network & 6 \\
\midrule
\multirow{3}{*}{Misconfigurations} & PyTorch & 4 \\
                                   & Communication & 6 \\
                                   & Dataloader & 5 \\
\midrule
\multicolumn{2}{l}{Low-efficiency code of users} & 45 \\
\bottomrule
\end{tabular}
\label{tab:issues}
\end{table}

\subsection{Case Study 1: Code-level Issues}
\label{subsec-case1}

\begin{figure}
\vspace{5.5pt}
	\centering
	\includegraphics[width=0.8\linewidth]{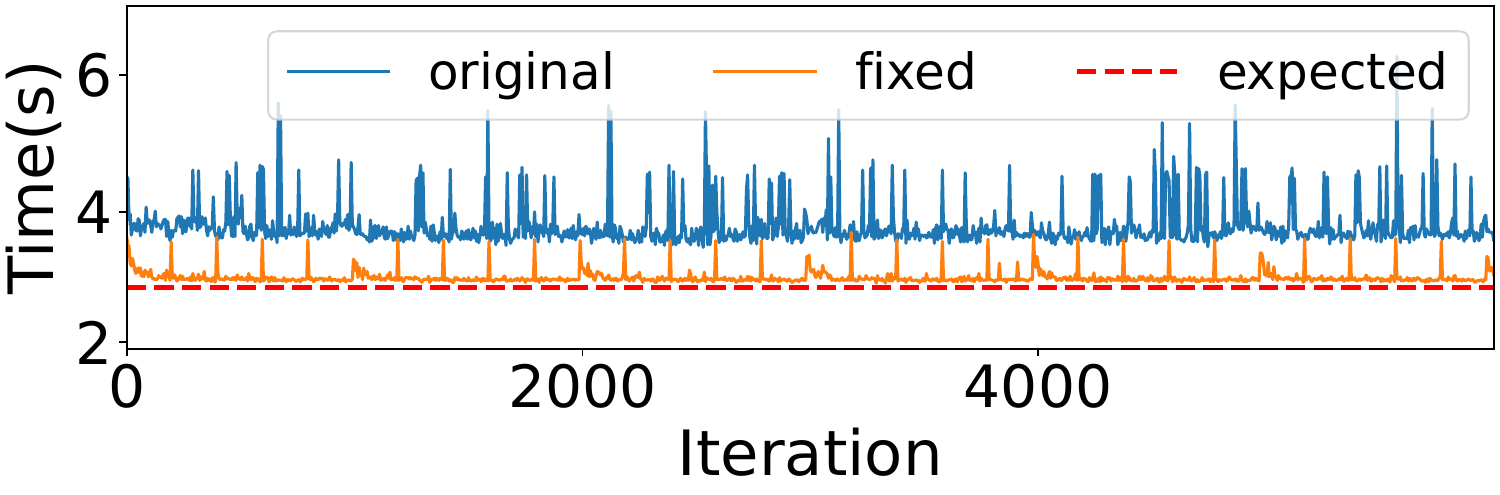}
 \vspace{-5pt}
	\caption{Iteration time of the \dlt in Case Study~1.}
	\label{fig:iteration cost case 2}
  \vspace{-5pt}
\end{figure}

A text-to-video \dlt job on 3,072 Nvidia H800 GPUs is expected to take 3.5 seconds per iteration, but is training at 5 seconds per iteration (see the "original" line in Figure~\ref{fig:iteration cost case 2}).

\begin{figure}[t]
	\centering
	\begin{subfigure}[t]{0.48\linewidth}
		\centering
		\includegraphics[width=\linewidth]{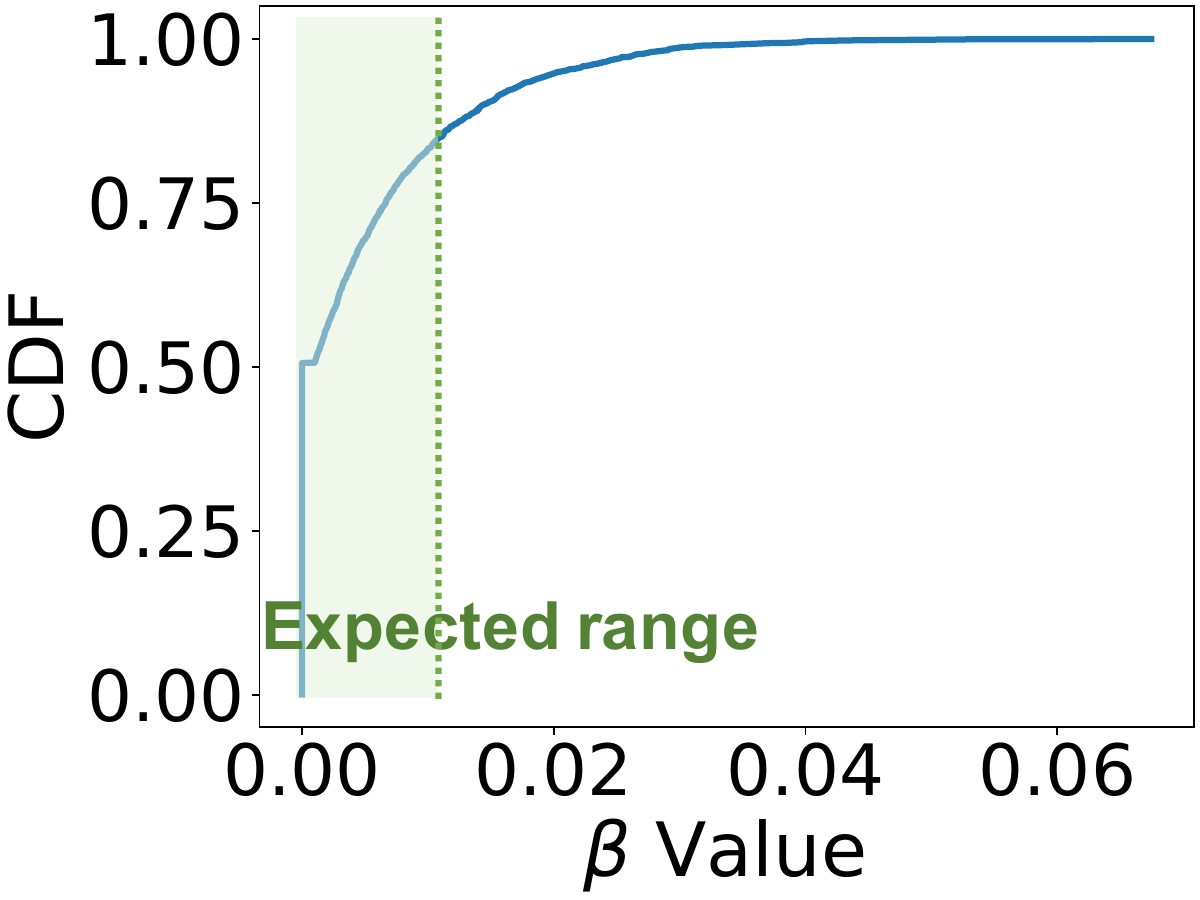}
		\caption{\texttt{recv\_into} of 3,072 workers.}
		\label{fig:recv_cdf}
	\end{subfigure}
	\hfill
	\begin{subfigure}[t]{0.48\linewidth}
		\centering
		\includegraphics[width=\linewidth]{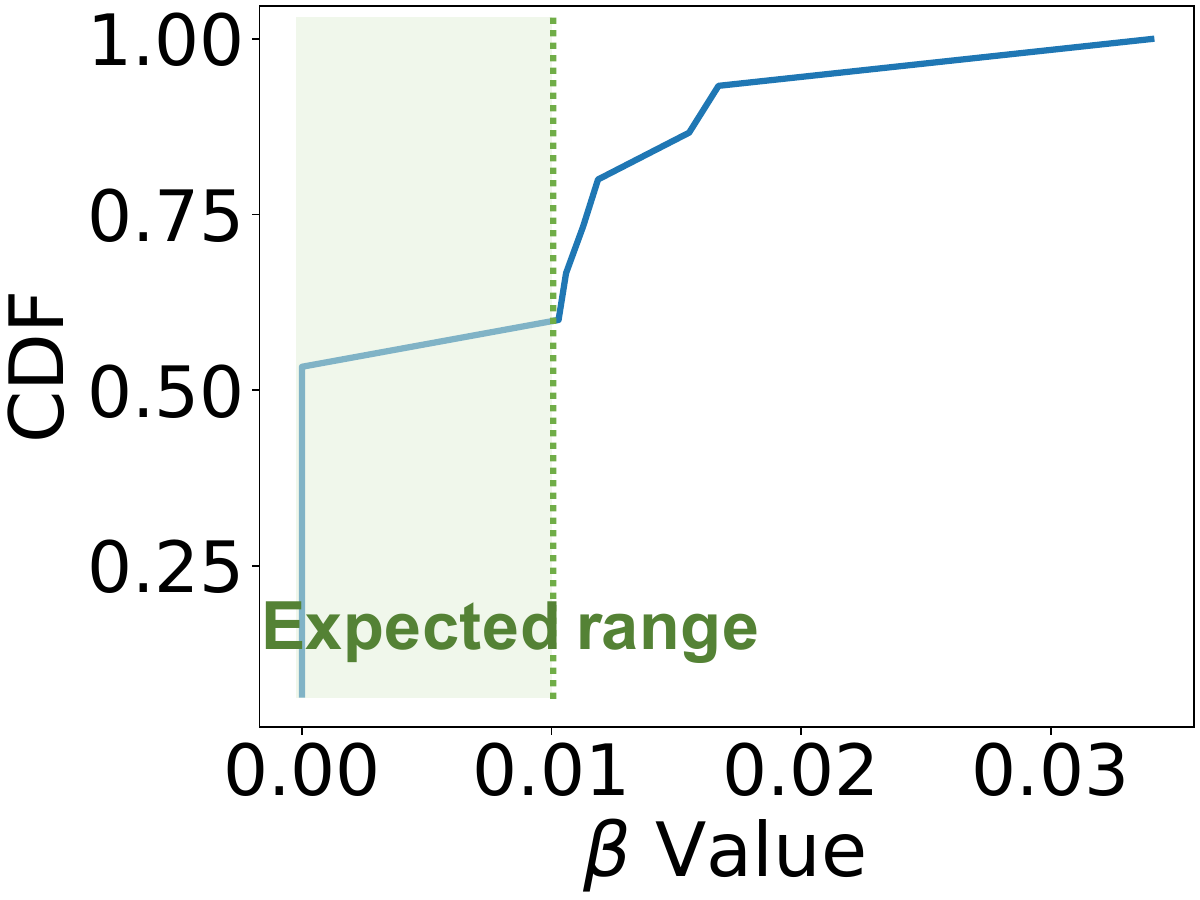}
		\caption{\texttt{forward} of 16 workers. The data of the rest 3,056 are destroyed by the cluster.}
		\label{fig:forward_cdf}
	\end{subfigure}
 \vspace{-10pt}
	\caption{CDF of $\beta$ values in two functions in Case Study~1.}
  \vspace{-10pt}
	\label{fig:cdf_case2}
\end{figure}

\para{Problem~1: Low socket throughput in data loader.}
\sys outputs abnormal function behaviors on the built-in function \texttt{\small recv\_into} of the socket object (a low-level function call in the data loader) on multiple workers. As a GPU-independent function, we expect that the \dlt is bottlenecked by it in no more than $1\%$ of the time ($\beta\le0.01$, as explained by ``expected range'' in \S\ref{subsec-design-analyzer}). However, as shown in Figure~\ref{fig:recv_cdf}, there are a lot of workers with a high value of $\beta$ for the \texttt{\small recv\_into} function (so they have a high value of $D_{f, w}$). This indicates low efficiency in data loading from the storage. 

\para{Problem~2: Inefficient implementation of Python function.}
Similar to the first problem, \sys outputs abnormal behaviors on the Python function \texttt{\small forward} with large $\beta$ values, shown in Figure~\ref{fig:forward_cdf}. This function performs CPU computation and launches GPU kernel functions. A large $\beta$ value indicates the function is bottlenecked by CPU computation, which needs to be optimized.

\para{Problem~3: Asynchronous garbage collection.}
Apart from the above functions, \sys also reports large $\beta$ values for some other Python functions like \texttt{\small gradmode.py:\_\_init\_\_()} and \texttt{\small \_flat\_param.py: \_get\_unflat\_views\_unaligned}. However, these functions are not CPU-intensive, and their executions with large $\beta$ are randomly distributed among all workers. 
This indicates these workers are executing Python garbage collection asynchronously. Each time a worker is collecting garbage, other workers have to wait for it, leading to GPUs being idle and reduced training throughput.

\para{Limitations of existing approaches.}
The customer did use Torch Profiler and Nsight System to profile their \dlt job offline.
As discussed in \S\ref{subsec-motiv-current}, profilers generate too much data; the common practice is profiling a few training iterations only in a fixed host (\eg, the rank-0 host). However, none of the three problems were detected, because in most cases the root cause of a low-performance user-level function is due to temporal pause on GPU/host memory allocation (\eg, garbage collection, GPU memory fragmentation, \etc) or access to remote storage. 
This only happens in a few random workers in each iteration, not all workers.
Consequently, the offending function is difficult to detect without profiling all workers.
In contrast, \sys has a global view of function behaviors of {\it all} workers to identify abnormal ones with high efficiency.

\para{Fixes.}
For the data loading problem, we helped the customer load the input data from our parallel file system rather than the legacy object storage service, which significantly accelerated data loading. 
For the Python garbage collection problem, the solution was to call the garbage collection every 200 iterations. This guaranteed all workers collect garbage at the same time, thus avoiding mutual waiting. Implementation optimization of the function \texttt{\small forward} is not trivial since it contains many CPU computation operations.
At present, the iteration time of the \dlt has reached $\sim$3.6 seconds, as shown in Figure~\ref{fig:iteration cost case 2}.

\vspace{-5pt}
\subsection{\edit{Case Study~2: Mixed Code-Hardware Issues}}
\label{subsec-case2}
\vspace{-2.5pt}

\begin{figure}
	\centering
	\includegraphics[width=0.95\linewidth]{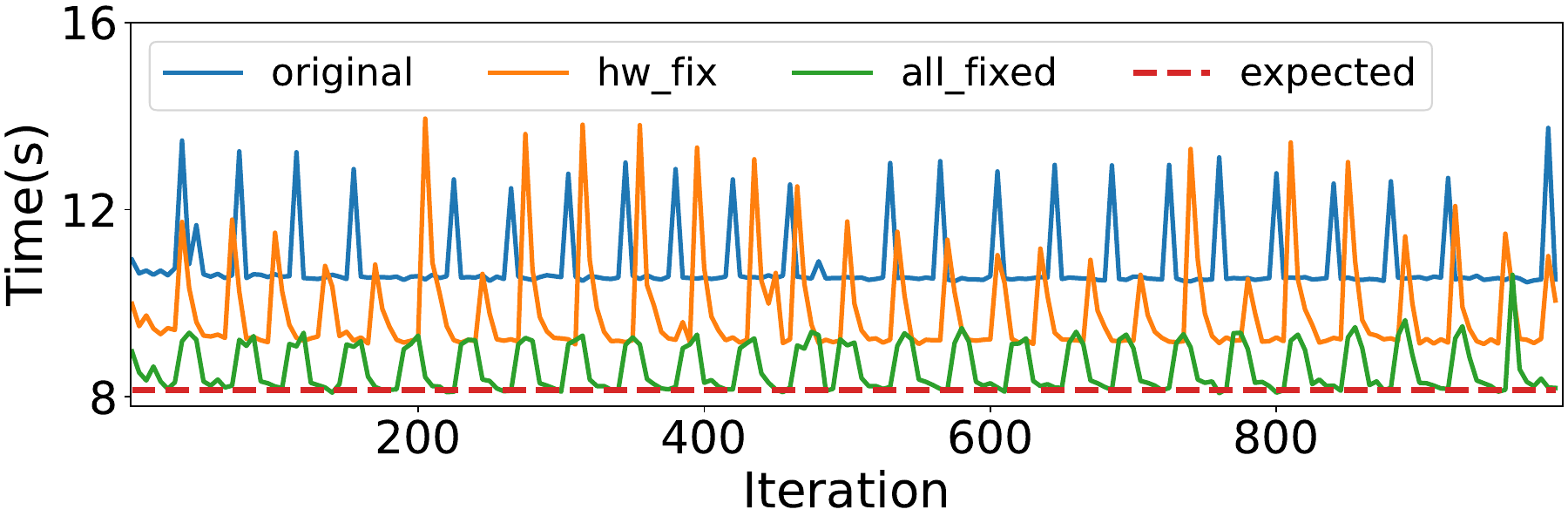}
 \vspace{-5pt}
	\caption{Iteration time of the \dlt in Case Study~2.}
 \vspace{-10pt}
	\label{fig:case_3_throughput}
\end{figure}

A video generation \dlt job on 3,400 Nvidia H800 GPUs is expected to take 8.5 seconds per iteration, but is training at 10.5 seconds per iteration ("expected" and "original" in Fig.~\ref{fig:case_3_throughput}). And the job crashes every few hours. 

\begin{figure*}
	\centering
	\begin{subfigure}[t]{0.24\linewidth}
		\centering
		\includegraphics[width=\linewidth]{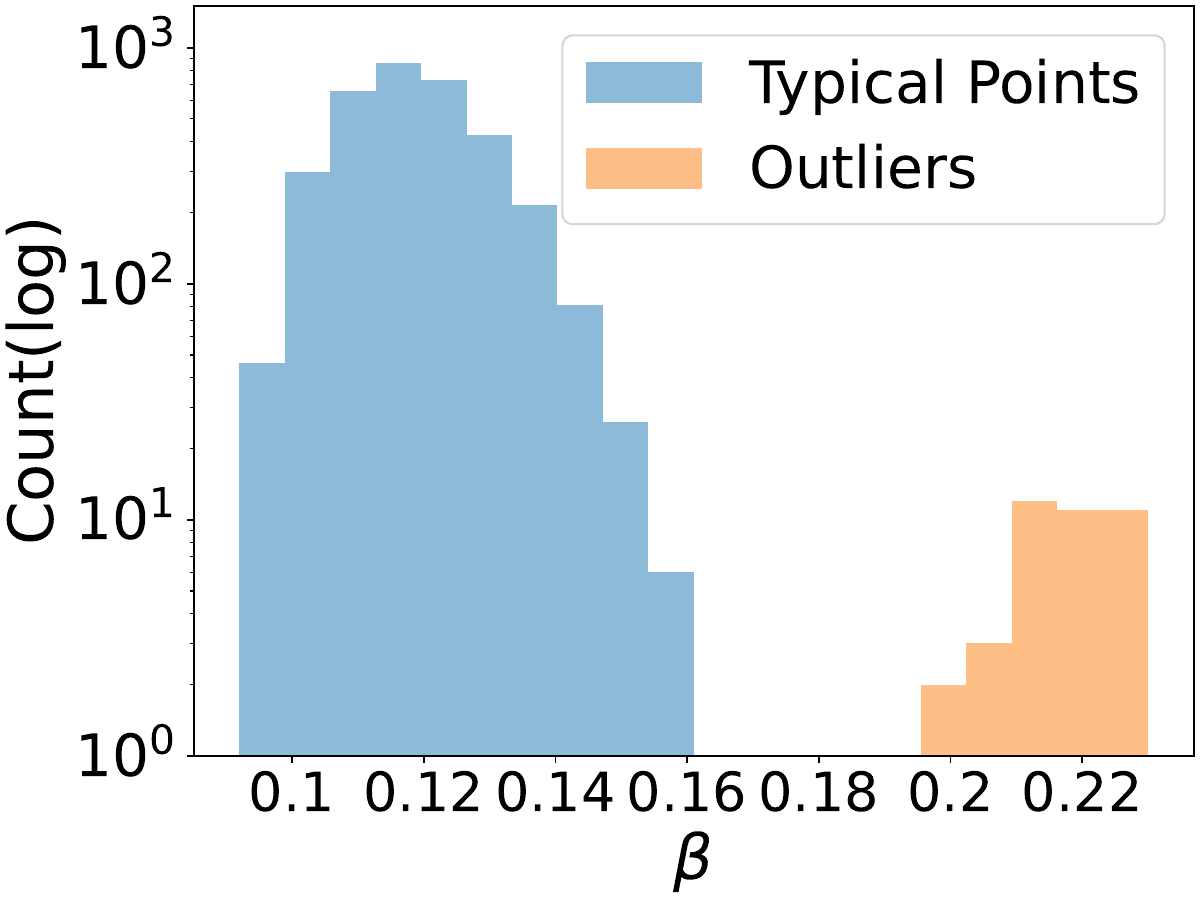}
		\caption{The $\beta$ value of \texttt{\small SendRecv} of 3,400 workers. This value is widely distributed across the 9\%-16\% range, with a small number (40 in total) of outliers distributed in 20\%-23\%.}
		\label{fig:case3_1}
	\end{subfigure}%
	\hfill
	\begin{subfigure}[t]{0.24\linewidth}
		\centering
		\includegraphics[width=\linewidth]{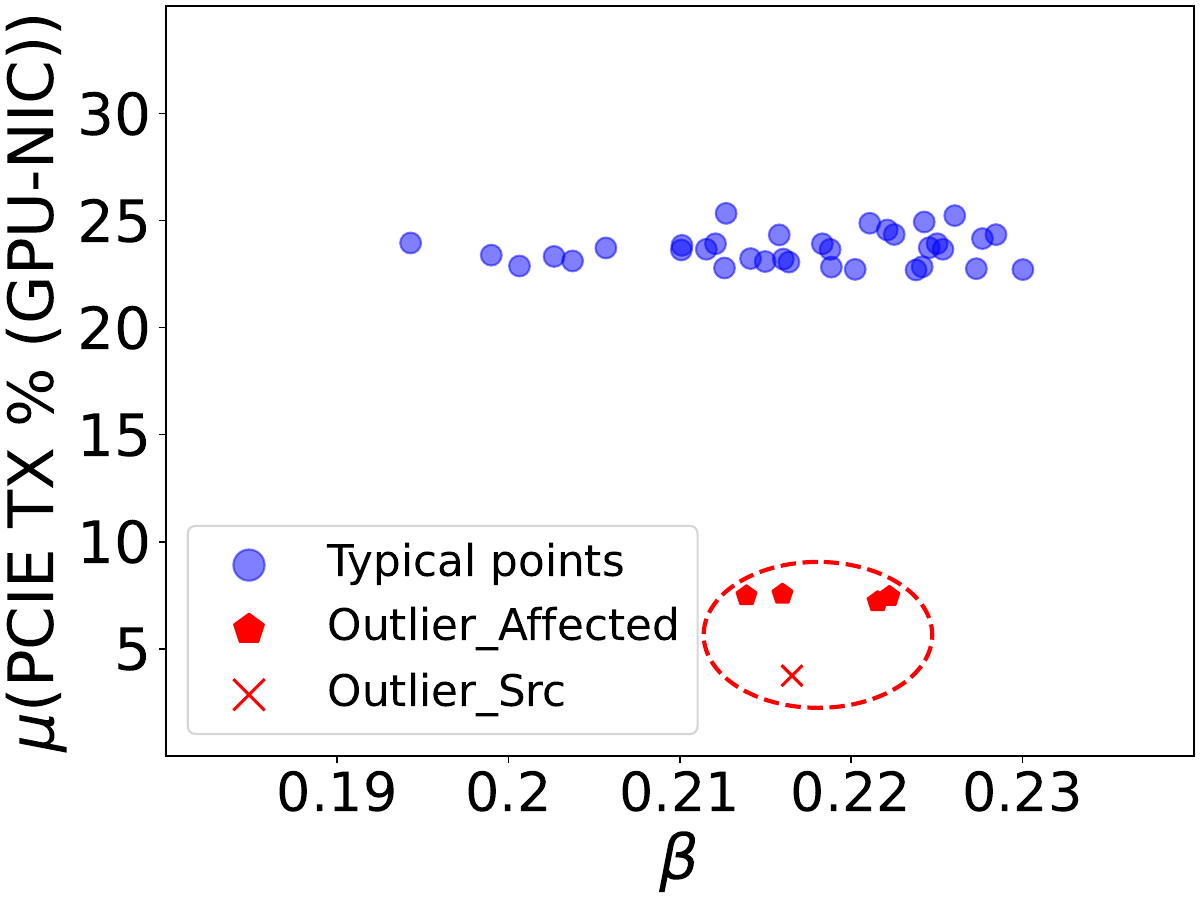}
		\caption{\texttt{\small SendRecv}'s $\beta$ and $\mu$ (GPU-NIC) of 40 workers with high $\beta$ (Outliers in Fig.~\ref{fig:case3_1}). Among the 40 workers, five have lower $\mu$ and one of them is significant lower which is the location of NIC down.}
		\label{fig:case3_2}
	\end{subfigure}%
    \hfill
	\begin{subfigure}[t]{0.24\linewidth}
		\centering
		\includegraphics[width=\linewidth]{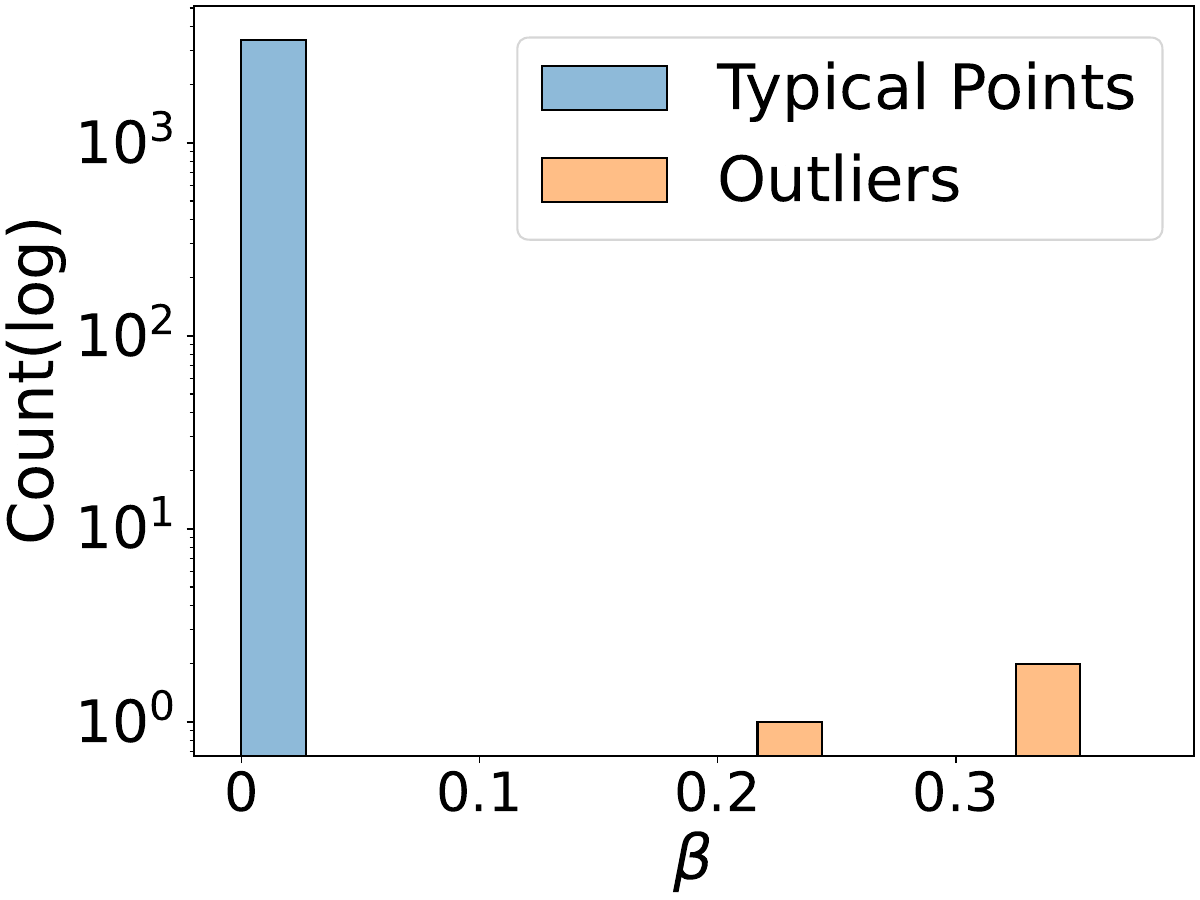}
		\caption{The $\beta$ value of \texttt{\small pin\_memory} of 3,400 workers. Most workers (3,397) is close to zero but three of them in range $23\%-33\%$.}
		\label{fig:case3_4}
	\end{subfigure}
	\hfill
	\begin{subfigure}[t]{0.24\linewidth}
		\centering
		\includegraphics[width=\linewidth]{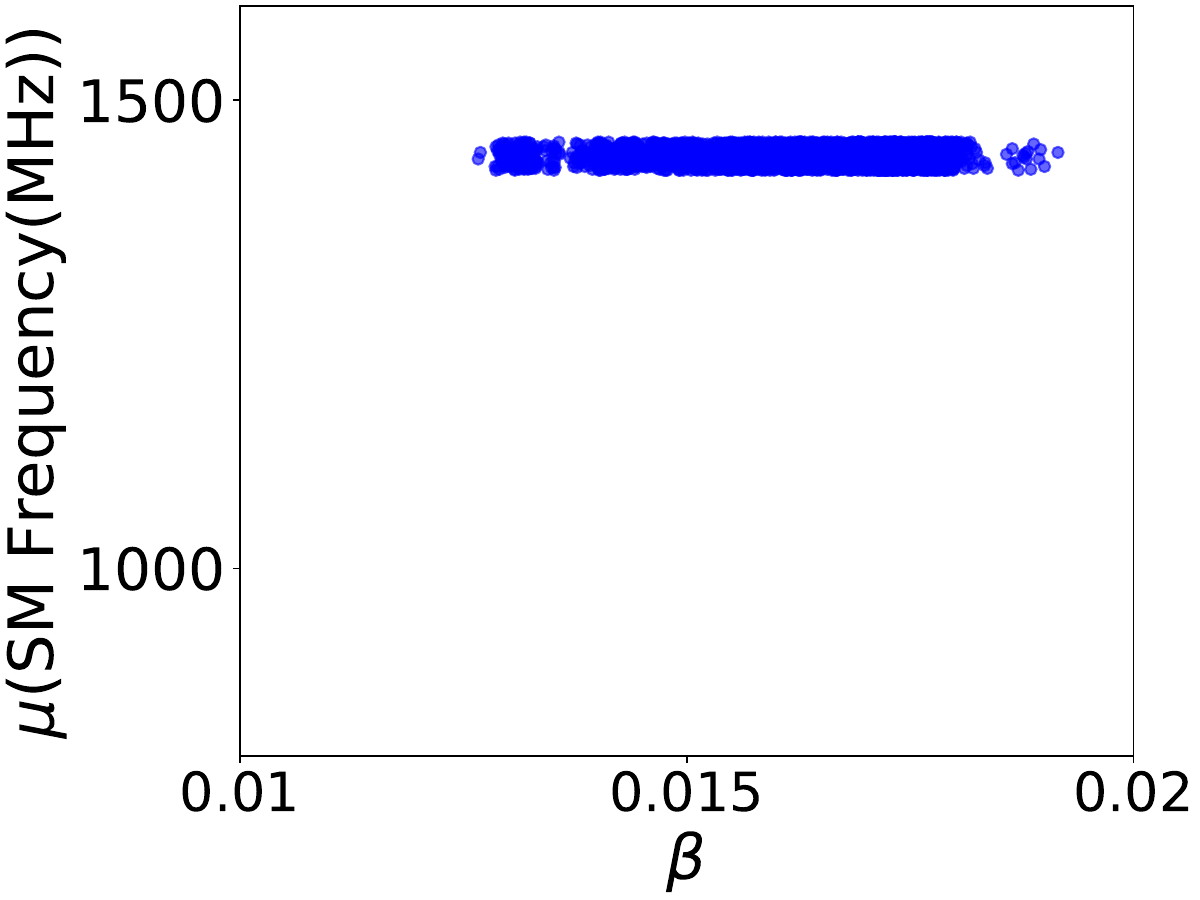}
		\caption{The $\beta$ and $\mu$ value of \texttt{\small chunk\_cat\_cuda\_kernel<float, c10::BFloat16>} of 3,400 workers. All workers have similar $\mu$ but different $\beta$, indicating the workload imbalance.}
		\label{fig:case3_3}
	\end{subfigure}
\vspace{-10pt}
	\caption{Function execution behaviors in the Case Study~2.}
\vspace{-10pt}
    \label{fig:case3}
\end{figure*}

\para{Problem~1: Low cluster network throughput.}
As shown in Fig.~\ref{fig:case3_1}, the $\beta$ value of most \texttt{\small SendRecv} functions ranges between 9\% and 16\%. However, our customer believes that the data amount in \texttt{\small SendRecv} is equal across all workers, so the $\beta$ values of \texttt{\small SendRecv} for all workers are expected to be identical. 
Moreover, based on the data amount in \texttt{\small SendRecv} and the NIC hardware, the $\beta$ value is expected to be $\sim$6\%---indicating much lower cluster network throughput than expected. The root cause is that affinity-based flow scheduling is not deployed on this cluster, so inter-host data flow is not optimized.

\para{Problem~2: NIC down.}
Fig.~\ref{fig:case3_1} shows another problem that the $\beta$ values of \texttt{\small SendRecv} of 40 workers (20\%-23\%) are much higher than most workers ($9\%-16\%$). Among the 40, one has high $\Delta_{f, w}$ due to the significantly lower $\mu$ than the other 39 (Fig.~\ref{fig:case3_2}), indicating a lower bandwidth between the GPU and the NIC.
Our network engineers confirmed that the NIC corresponding to the worker is down.

\para{Problem~3: Pin memory in ultra-high frequency.}
Fig.~\ref{fig:case3_4} shows that three random workers spent at most one third of time on \texttt{\small pin\_memory} ($23\%-33\%$ $\beta$ value). Although this does not happen on the other 3,397 workers, they have to wait for the three workers to finish one iteration together, leading to reduced training throughput. Our customer confirmed that \texttt{\small pin\_memory} operations come from \texttt{\small data\_loader} processes.

\para{Problem~4: Load imbalance.}
\sys finds that almost all GPU computation kernel functions have similar $\mu$ but significantly different $\beta$ across workers. Fig.~\ref{fig:case3_3} shows one GPU kernel function for example. The busiest GPU spends 46\% more time computing than the most idle one. This indicates that although workers execute GPU kernels with equal performance, some workers launch much more GPU kernels than others, leading to load imbalance. Our customer confirmed that the model is a video-to-video model, for each training iteration, video inputs for workers have different duration, requiring different amount of GPU kernel launches. Workers with less GPU computation have to wait for others with more GPU computation to finish one iteration together; this leads to poor training efficiency (\eg, low Model FLOPs Utilization).

\para{Limitations of existing approaches.}
For Problem 1, hardware monitoring on the cluster network showed no warning. The problem was low-efficiency flow scheduling that failed to fully utilize the network. 
For Problem~2, the affected host was newly added to the cluster, and the monitoring agent had not been updated, therefore, failed to alarm the problem.
For Problem~3, the pure code-level problem, it only happened on three of 3,400 workers in one iteration, so it was hard to be captured offline (as explained in \S\ref{subsec-case1}).
For Problem~4, although a GPU may receive a long video in one iteration to reach high utilization, due to the randomness of input scheduling, it does not happen in all iterations. Based on our GPU utilization monitor, in general all GPUs have similar utilization in a period of time, so it failed to detect the problem.

\para{Fixes.}
We removed 20 hosts with the least network throughput in Fig.~\ref{fig:case3_1} to mitigate Problem~1, including the host with NIC down (Problem~2), then the iteration time reduced from 10.5s to 9.5s ("hw\_fix" line in Fig.~\ref{fig:case_3_throughput}). 
For Problem~3, our customer decreased the number of \texttt{\small data\_loader} processes to reduce the memory overload. It not only decreased the iteration time (from 9.5s to 9.2s), but also solved the crash problem.
For Problem~4, we helped the customer balance the input between workers. Currently, this \dlt is training at 8.5 seconds per iteration (the "all\_fixed" line in Fig.~\ref{fig:case_3_throughput}) without crashes. With all of the above efforts, the input data processing rate was improved from 7,173 samples/day to 9,644 samples/day, a 34\% end-to-end performance improvement.

\subsection{\edit{Case Study~3: Diagnose and auto-fix customer-code bugs with AI support}}
\label{subsec-case5}
A robotics (embodied AI) model's training of 128 GPUs got stuck for a long time, and the customer requested a diagnosis.

\para{Problem detected: Stucked in dataset preloading.} In this stalled training job, \sys found a single worker behaving differently from the rest. On this worker, a data-loading/preprocessing thread is blocked in \texttt{\small queue.put()} inside \texttt{\small dynamic\_robot\_dataset.\_preload()} (waiting on a Python thread lock), indicating the input pipeline is stuck/back-pressured. Meanwhile, other workers are either sleeping in dataset management routines (\eg, \texttt{\small \_monitor\_config}, \texttt{\small \_run\_threads}) or waiting in JAX~\cite{jax} execution, suggesting they are idle and effectively waiting for this worker to make progress. Overall, the hang is most consistent with a Python-side data pipeline deadlock or queue blockage in the dataset prefetch/preload logic.

\para{Automatically fixed by AI.} We instructed the customer to use \sys's output (the stuck Python function) together with the relevant code as a prompt to Cursor~\cite{cursor} (Sonnet 4.5), an AI coding tool. Cursor immediately identified the bug: during data fetching, a logging/debug print accessed array[0] on a sharded distributed array, implicitly triggering an \texttt{\small all-gather}. Since this collective was invoked at an unexpected point (not executed by all ranks), it led to a distributed deadlock and stalled the training job. Cursor then automatically generated the fix for the buggy code, and the training job resumed normally.

\vspace{-4.5pt}
\subsection{System Overhead}
\label{subsec-overhead}

\begin{figure}[!t]
	\centering
	\includegraphics[width=\linewidth]{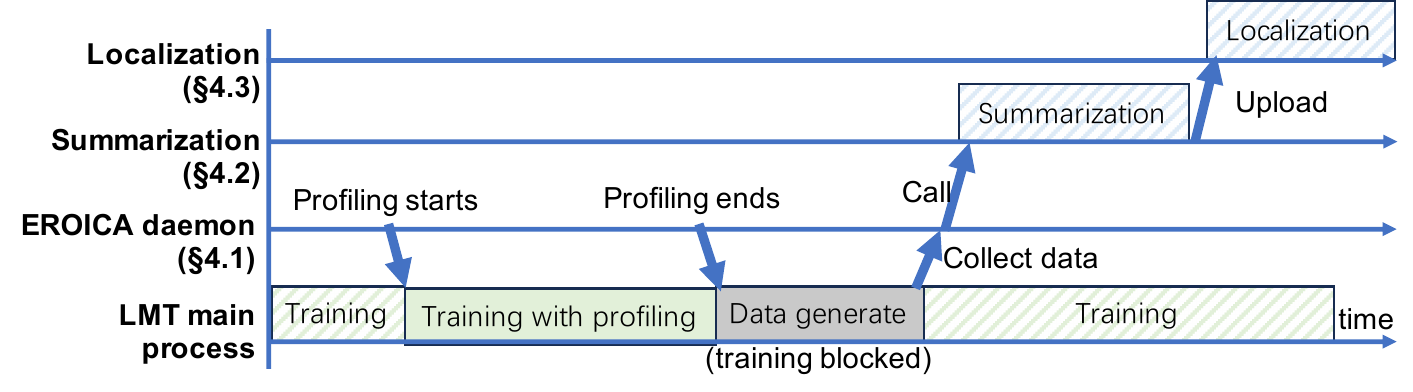}
        \vspace{-15pt}
	\caption{\sys's overhead in timeline view.}
  \vspace{-5pt}
	\label{fig:overload}
\end{figure}

Figure~\ref{fig:overload} illustrates \sys's overhead. When \sys starts profiling, the \dlt is executed with high-precision profiling for a 20-second window (\S\ref{subsec-design-collector}). When the profiling window ends, \dlt will be blocked by data generation for a short period. After that, \dlt continues running with no extra overhead, as the behavior pattern summarization is done in a separate process (using a different CPU core), and the root cause localization is performed remotely from the worker.

\begin{figure}
	\centering
	\begin{subfigure}[t]{0.32\linewidth}
		\centering
		\includegraphics[width=\linewidth]{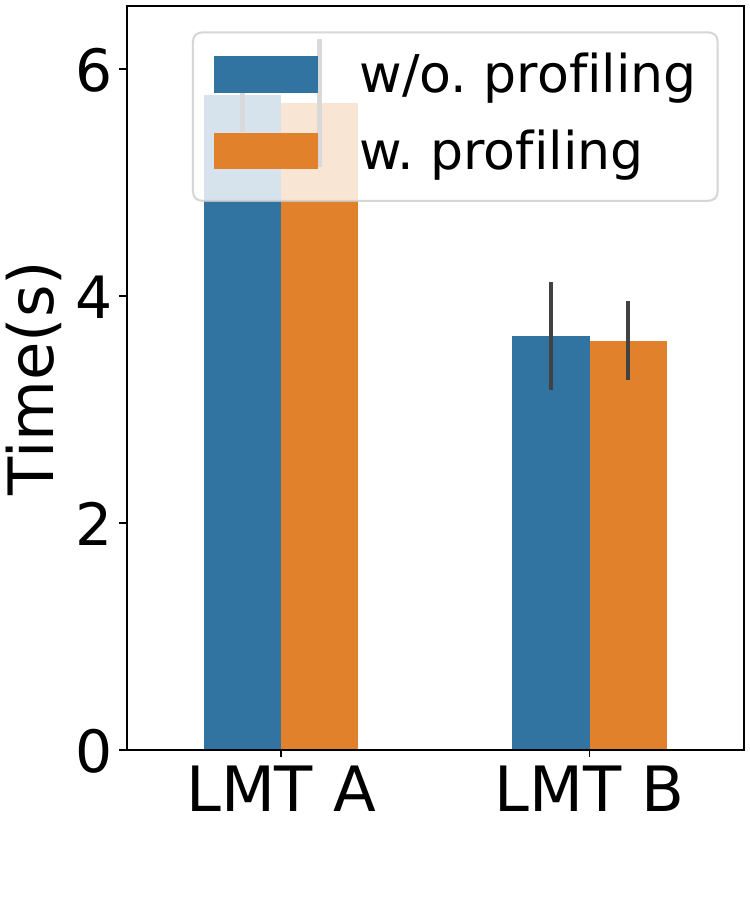}
		\caption{\dlt{} iteration time with and without profiling.}
		\label{fig:overhead}
	\end{subfigure}
	\hfill
	\begin{subfigure}[t]{0.32\linewidth}
		\centering
		\includegraphics[width=\linewidth]{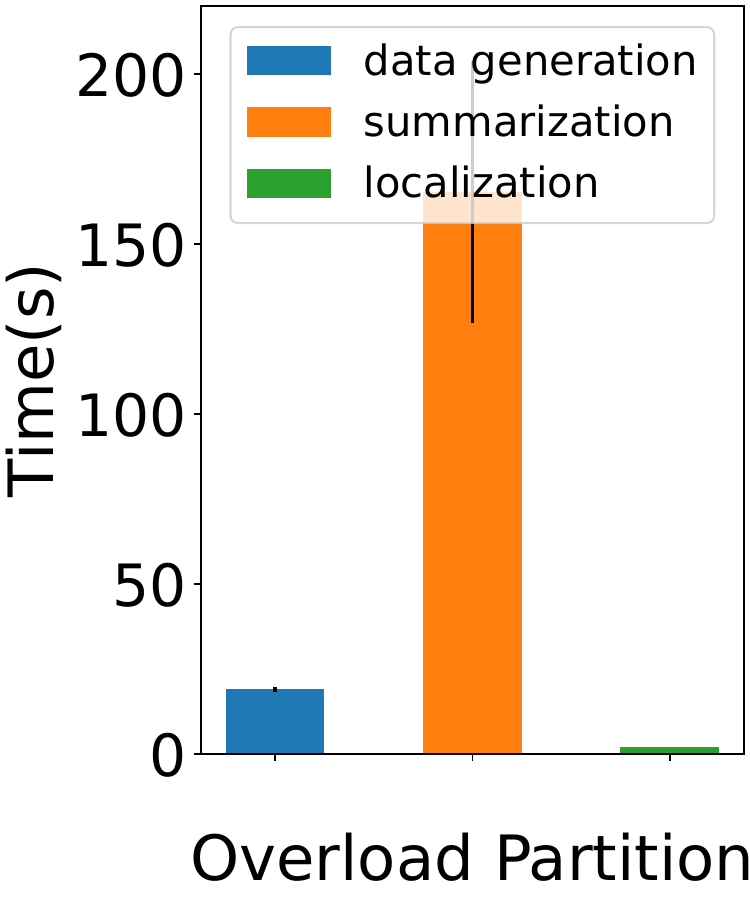}
		\caption{Time spent on different components.}
		\label{fig:profile_partition_dltb}
	\end{subfigure}
    \hfill
    \begin{subfigure}[t]{0.32\linewidth}
        \centering
        \includegraphics[width=\linewidth]{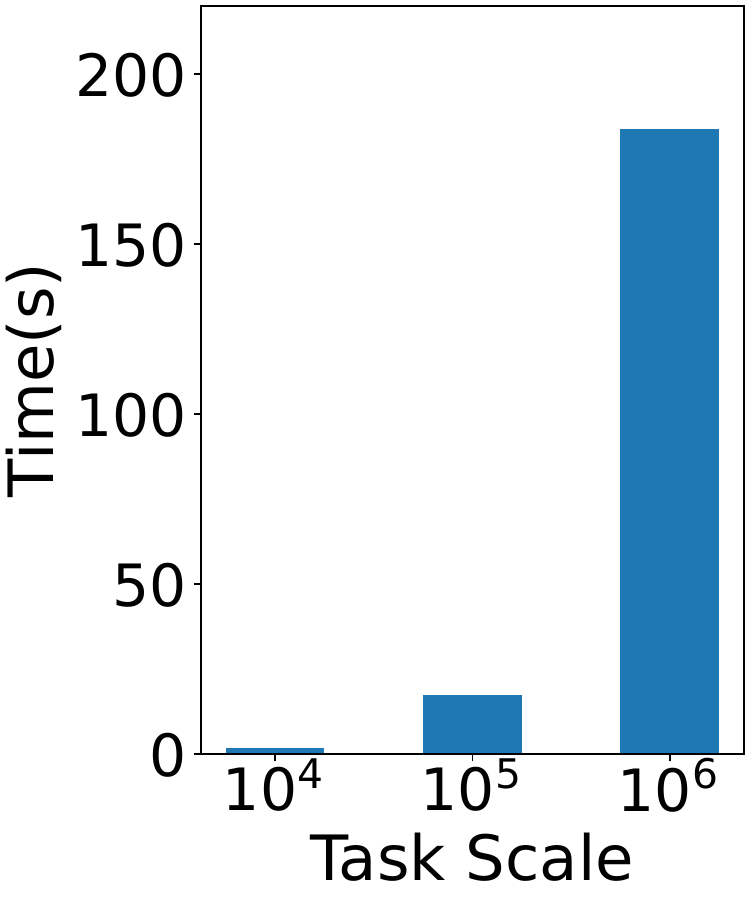}
        \caption{Time spent on localization under larger \dlt scales.}
        \label{fig:analysis overhead}
    \end{subfigure}
 \vspace{-5pt}
	\caption{The overhead introduced by \sys{}.}
 \vspace{-5pt}
	\label{fig:job_overhead}
\end{figure}

\para{Overhead on production \dlts.}
We evaluate two \dlts: \dlt-A of 3,072 GPUs in \S\ref{subsec-case1} and \dlt-B of 3,400 GPUs in \S\ref{subsec-case2}.
Figure~\ref{fig:overhead} shows the average iteration time during training, with and without profiling. 
Figure~\ref{fig:profile_partition_dltb} shows the duration of data generation, pattern summarization‌, and localization.
The results show that profiling does not affect the performance of production \dlts; summarization and localization complete within three minutes. The main overhead on \dlt is data generation ($\sim$20) seconds, which is \camera{incurred just once and} accepted by all our customers.
\camera{Notably, pattern summarization, uploading, and root cause localization occur outside the training process. 
Although these steps may last for minutes, they introduce zero overhead to the training task.
Regarding memory usage, although profiling consumes tens of GBs on each host, this is negligible given that production training hosts typically have 1--2~TB of host memory.}

\edit{We also evaluate the overhead with different models and configurations in Appendix~\ref{overhead_with_config} to show \sys is able to make diagnosis with high efficiency in diverse situations.}


\para{Scalability to 1,000,000-GPUs \dlts.}
In \sys, profiling, data generation and pattern summarization are all distributed in each worker's container and executed in parallel, so their overhead is independent of the \dlt scale. The only overhead proportional to the \dlt scale is root-cause localization, which is executed on a single CPU core.
We generate simulated \patternss as input, to evaluate the overhead of localization for larger-scale \dlts. 
Figure~\ref{fig:analysis overhead} shows that the localization requires only three minutes for a 1,000,000-GPU \dlt. Together with the time spent on profiling, data generation, and pattern summarization, \sys can make an end-to-end performance analysis for a 1,000,000-GPU \dlt 
within 7 minutes. 

\vspace{-5pt}
\section{Discussions: Enhancing \sys with AIOps}
\label{limitations}
\vspace{-2.5pt}

\sys does not guarantee automatic identification of the root causes for all possible performance issues.
(1) For most performance issues, \sys's output of the problematic functions and their corresponding \patternss is sufficient to identify the root cause or can be fed into AI to determine the root cause (like the case in \S\ref{subsec-case5}). However, some user-defined Python functions and CPU kernels have complex code and logic. Therefore, even when \sys identifies that these functions have problems (\eg, excessively long execution times), analyzing the root causes may still require manual inspection of their code.
\camera{(2) \sys does not ensure detecting problems outside the training task. For example, a background process running on the physical host may incidentally consume too many resources, leading to a performance issue for the training task in the container on that host. A case study is presented in Appendix~\ref{subsec-case4}.}

\camera{To bridge the last-mile gap between abnormal function behavior and the root cause, we are instructing our customers to combine \sys's output with additional information (the code of the function with abnormal behavior, background processes with system call activities, and hardware configuration/utilization) and generate a standardized prompt for an AI model to diagnose the root cause. In production, while the AI provides correct diagnoses only in a subset of cases, it yields useful hints in most cases, which is particularly valuable for performance issues caused by user code. Therefore, in our research on AIOps (Artificial Intelligence for IT operations) for \dlt, \sys's output serves as the core component for prompt construction.}

\vspace{-10pt}
\section{Related Work}
\vspace{-5pt}

\com{{\bf Hardware monitoring tools.}
GPUs can be monitored by a series of tools developed by NVIDIA.
NVML~\cite{nvml} is a comprehensive, C-based API library. It provides a standardized interface for users to interact with the GPUs, allowing them to actively sample the status of these devices. DCGM~\cite{dcgm} is specifically designed for monitoring GPUs in large data centers. It offers a centralized and scalable monitoring method, making it highly suitable for enterprise environments with extensive GPU deployments. 
CPUs can be monitored by PCM~\cite{pcm}, an application programming interface for monitoring performance and energy metrics of Intel processors. 
For monitoring NICs (\eg, Mellanox), mstflint~\cite{mstflint} is developed to access packet or byte-level throughput data.

\para{\dlt profiling and analysis.}
Nsight Systems and Nsight Compute provide comprehensive performance analysis and profiling capabilities. Nsight Systems offers a holistic view of CPU, GPU, and memory usage, while Nsight Compute focuses on detailed kernel performance metrics for CUDA applications, helping developers identify and optimize computational bottlenecks.
Torch Profiler is a tool within the PyTorch ecosystem that assists in optimizing PyTorch models by providing detailed insights for each function.}

{\bf Hardware monitoring tools.}
Accelerators expose rich telemetry through vendor tools. For NVIDIA GPUs, NVML~\cite{nvml} provides a C-based API to query and sample device status (e.g., utilization, temperature, power, memory, and error states). DCGM~\cite{dcgm} targets large clusters by offering centralized, scalable GPU health/performance monitoring and diagnostics for fleet management. CPU performance and energy metrics can be collected via Intel PCM~\cite{pcm} (e.g., cache and memory bandwidth statistics). For NICs (e.g., Mellanox), \texttt{mstflint}~\cite{mstflint} provides access to device counters such as packet/byte throughput and link status.

\para{\dlt profiling and analysis.}
Nsight Systems and Nsight Compute provide complementary profiling for CUDA workloads. Nsight Systems captures an end-to-end timeline across CPU threads, GPU work, and memory transfers, whereas Nsight Compute focuses on per-kernel metrics (e.g., occupancy and memory throughput). Torch Profiler, integrated into PyTorch, reports fine-grained per-operator/per-function events across Python, CPU, and CUDA to attribute training time and identify bottlenecks.

\vspace{-7.5pt}
\section{Conclusion}
\vspace{-5pt}

This paper presents \sys, an online performance troubleshooting system for \dlt. \sys presents the process of profiling on performance degradation, behavior pattern summarization, and performance issue localization, which utilize \patternss of functions to make diagnosis. We deploy \sys in real-world production \clusters and solve sophisticated performance issues for real users.
\vspace{-7.5pt}
\section*{Acknowledgements}
\vspace{-10pt}

We thank our shepherd Changhoon Kim, and the anonymous
reviewers for their insightful comments.
Ennan Zhai is the corresponding
author.

\bibliographystyle{acm}
\bibliography{bibfile.bib}

\appendix
\newpage
\section*{\centering{APPENDIX}}
\label{sec:appendix}

\section{\edit{Case Study~4: Hardware Issues}}
\label{subsec-case3}

\begin{figure}[H]
\vspace{10pt}
	\centering
	\includegraphics[width=0.9\linewidth]{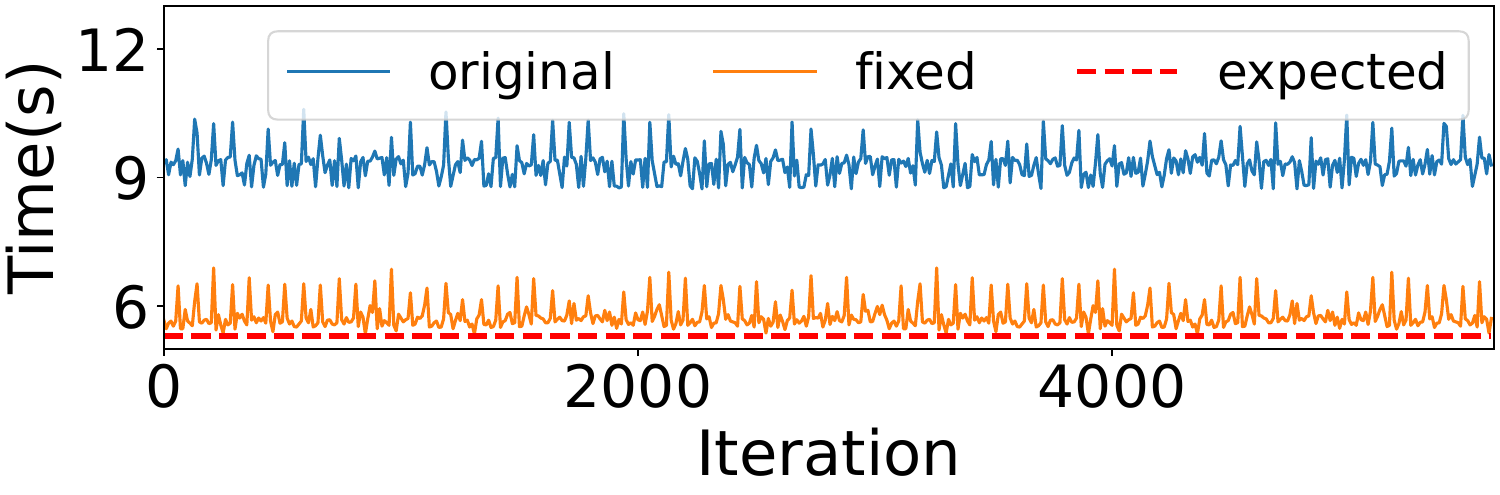}
	\caption{Iteration time of the \dlt in Case Study~4.}
	\label{fig:case_1_throughput}
\end{figure}

A text-to-picture \dlt job of 2,560 Nvidia H800 GPUs is expected to take 5 seconds per iteration, but is training at 9 seconds per iteration ("expected" and "original" in Fig.~\ref{fig:case_1_throughput}).

\begin{figure*}
	\centering
	\begin{subfigure}[t]{0.32\linewidth}
		\centering
		\includegraphics[width=0.7\linewidth]{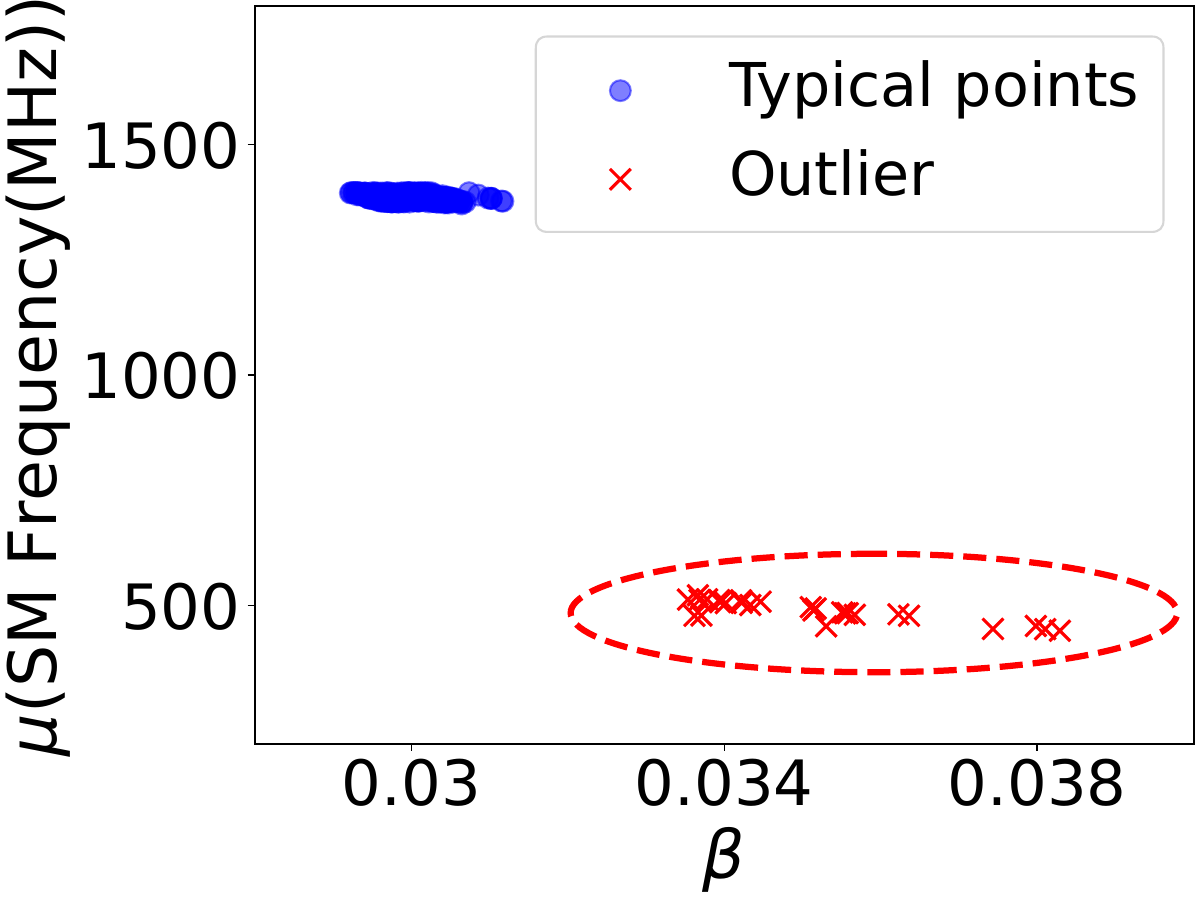}
		\caption{The value of $\beta$ and $\mu$ (average GPU SM frequency) of \texttt{\small GEMM} on 2,560 workers.}
		\label{fig:down gpu}
	\end{subfigure}%
	\hfill
	\begin{subfigure}[t]{0.32\linewidth}
		\centering
		\includegraphics[width=0.7\linewidth]{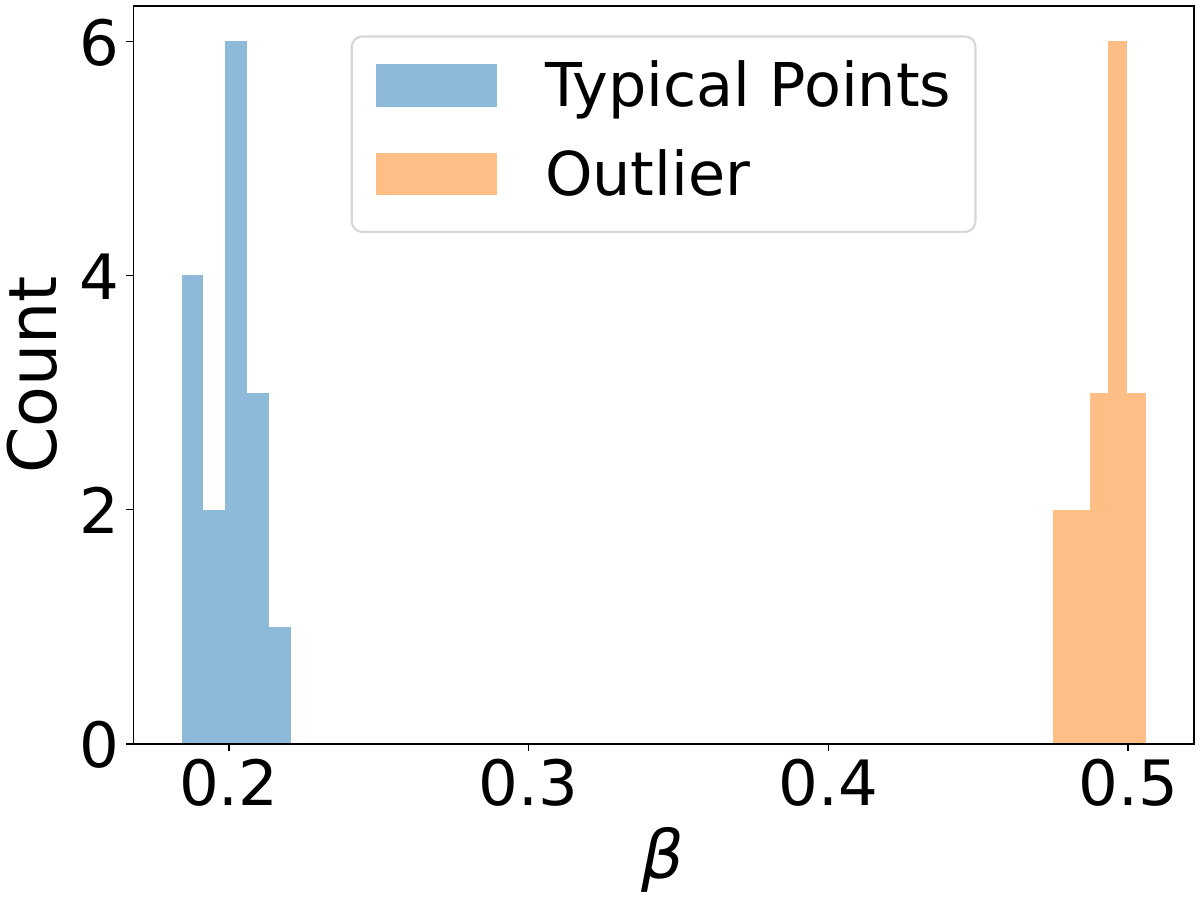}
		\caption{$\beta$ values of \texttt{\small AllGather\_RING}. There are 2,512 typical points and 48 outliers. We sampled 16 for both of them.}
		\label{fig:slow nvlink}
	\end{subfigure}%
	\hfill
	\begin{subfigure}[t]{0.32\linewidth}
		\centering
		\includegraphics[width=0.7\linewidth]{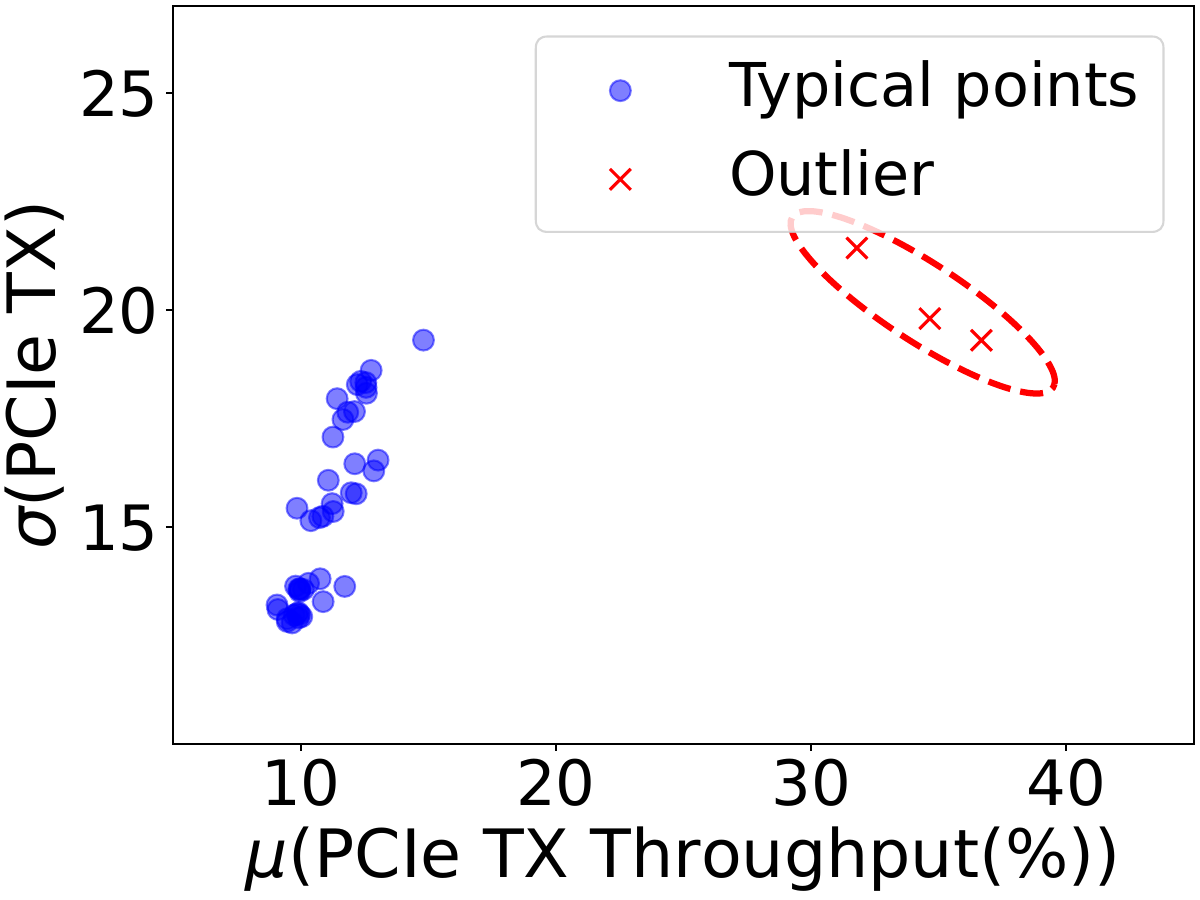}
		\caption{The value of $\mu$ and $\sigma$ (PCIe TX throughput from GPU to NIC) of \texttt{\small AllGather\_RING} on the 48 workers with high $\beta$ value.}
		\label{fig:down nvlink}
	\end{subfigure}
\vspace{-10pt}
	\caption{Function execution behaviors of \texttt{\small GEMM} and \texttt{\small AllGather\_RING} in Case Study~3. Each point represents the \pattern of one worker. Outliers are regarded as abnormal function executions with high $\Delta_{f, w}$ (\S\ref{subsec-design-analyzer}) and are output by \sys.}
    \label{fig:case1}
\end{figure*}

\para{Problem~1: GPU throttling.}
\sys outputs 300+ workers (located in 40+ hosts) with abnormal execution behaviors on multiple GPU computation kernel functions. Figure~\ref{fig:case1}\subref{fig:down gpu} shows the example of \texttt{\small GEMM} (one of the GPU computation kernel functions). GPU computation kernels on these workers have much larger $\beta$ and smaller $\mu$ than most workers, so they have high values of $\Delta_{f, w}$. This indicates that the GPUs of these workers are much slower, so they need a longer time for computation. 
Interestingly, \sys conducted multiple profiles of this \dlt, and in each profile, the slow GPU workers were not consistent, but they are intensively distributed among workers in certain racks, rather than evenly distributed among all workers. This indicates that GPUs from specific deployment batches have intermittent performance issues.
Hardware checks revealed recent \texttt{\small GPU\_NVSMI\_HW\_THROTTLE} alerts on some of these machines, confirming the risk of intermittent GPU throttling. We have contacted the hardware vendor for repairs on these machines.

\para{Problem~2: NVLink down.}
\sys also outputs 3 workers with abnormal execution behaviors on multiple collective communication functions. Taking \texttt{\small AllGather\_RING} (one of the collective communication functions) as an example, 48 workers (belonging to three data-parallel groups, where each group has 16 workers) have significantly larger values of $\beta$ than the other 2,512 (Figure~\ref{fig:slow nvlink}). Among these 48 workers, three of them have significantly larger $\mu$ than the other 45 (Figure~\ref{fig:case1}\subref{fig:down nvlink}), indicating a higher bandwidth usage of PCIe.
Hardware checks revealed NVLink Not Supported (NS) error on these three workers, which means all data flow sent to/from these three workers has to go through PCIe rather than NVLink. Since PCIe is much slower than NVLink, it leads to communication inefficiency.

\para{Limitations of hardware monitors.}
Section~\ref{subsec-motiv-current} discusses limitations of hardware monitoring. In this case, for Problem 1, the affected GPU throttling events occur at low frequency with short durations (100us–10ms), failing to trigger Nvidia's built-in alerts. The monitoring system, operates at minute-level granularity (with second-level precision in short intervals) to be light-weighted, also failed to capture such fine-grained performance fluctuations. For Problem 2, the affected machines were newly added to the cluster, and the monitoring agent had not been updated, thus failing to detect the problem.

\para{Fixes.}
We replaced the problematic hosts with standby hosts and restarted the \dlt. Figure~\ref{fig:case_1_throughput} shows the iteration time after the replacement (the "fixed" line).

\section{Case Study~5: Code Issue (\sys Failed to diagnose)}
\label{subsec-case4}
A reinforcement learning \dlt job of 8 Nvidia H800 GPUs (located at one host) is expected to take $\sim22$ seconds per training iteration (about hundreds of commits before the current version, called Version A), but is training at $\sim26$ seconds per training iteration (called version B).

\begin{figure}[t]
	\centering
	\begin{subfigure}[t]{0.48\linewidth}
		\centering
		\includegraphics[width=\linewidth]{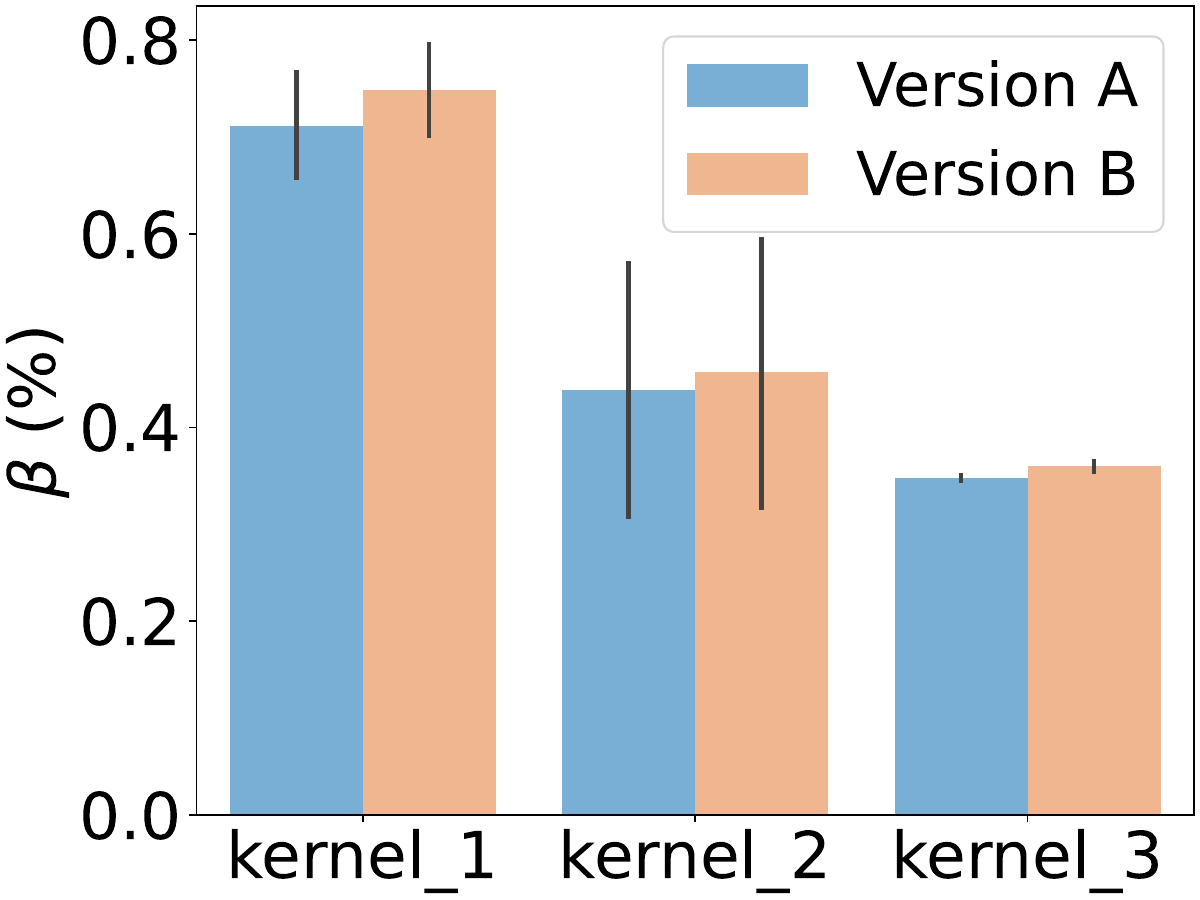}
		\caption{$\beta$ value comparison of three representative GPU computation kernel functions between Version A and B.}
		\label{fig:case4_1}
	\end{subfigure}
	\hfill
	\begin{subfigure}[t]{0.48\linewidth}
		\centering
		\includegraphics[width=\linewidth]{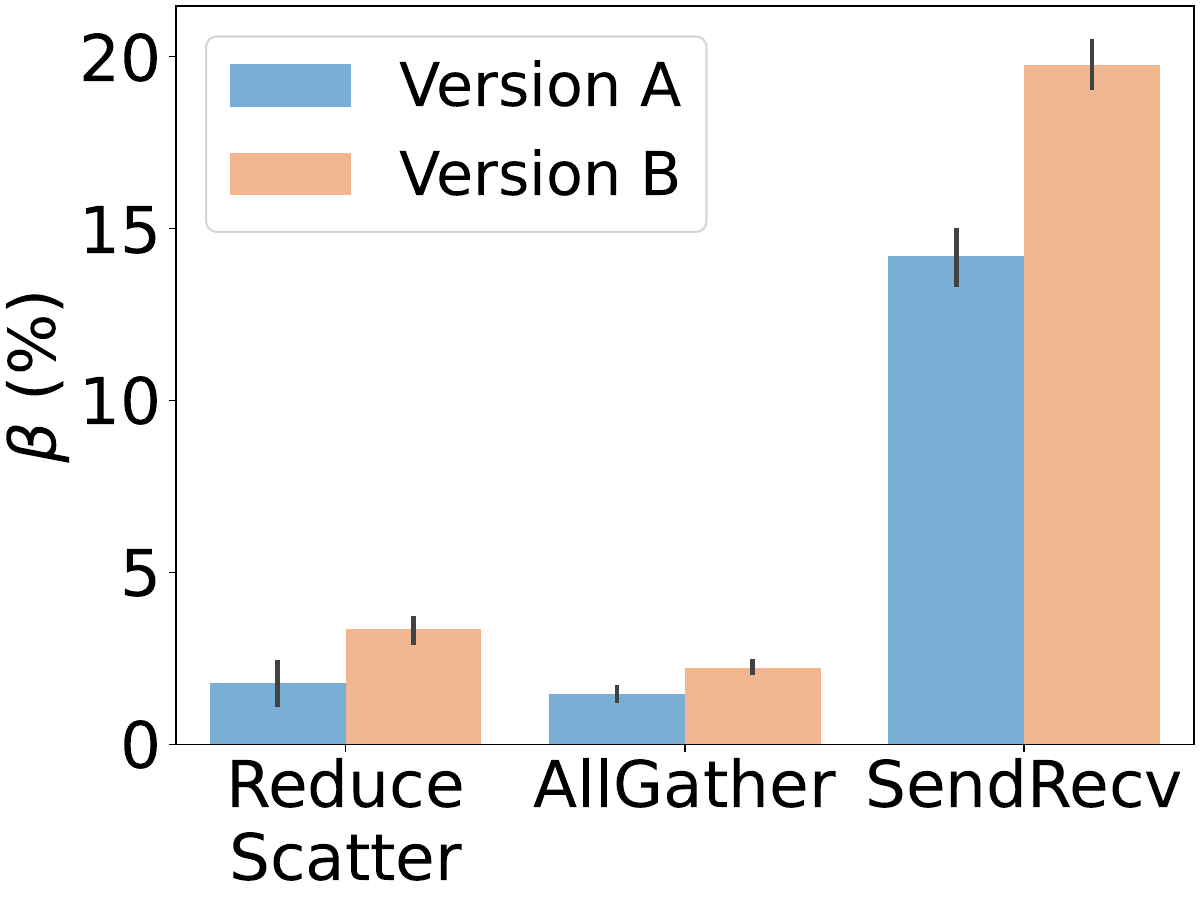}
		\caption{$\beta$ value comparison of representative collective communication functions between Version A and B.}
		\label{fig:case4_2}
	\end{subfigure}
 \vspace{-10pt}
	\caption{$\beta$ values of GPU computation and communication functions in Case Study~4 (we omit kernel names of GPU computation kernel functions since they are too long).}
  \vspace{-10pt}
	\label{fig:case4}
\end{figure}

\begin{figure*}[!t]
    \centering
    \includegraphics[width=\textwidth]{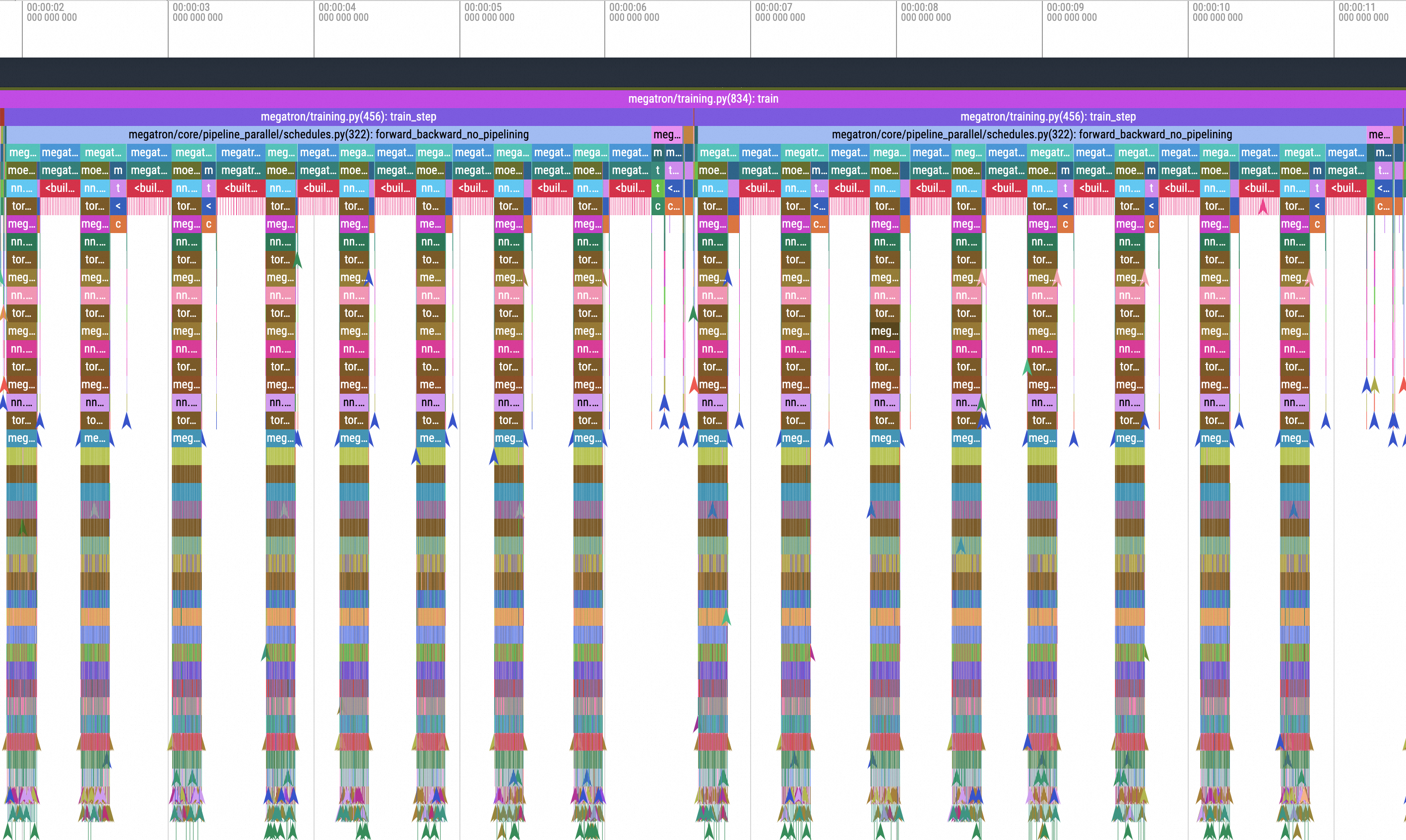}
    \caption{The runtime behavior of an MoE training.}
    \label{fig:perfetto_moe}
\end{figure*}

\begin{figure*}[!t]
    \centering
    \includegraphics[width=\textwidth]{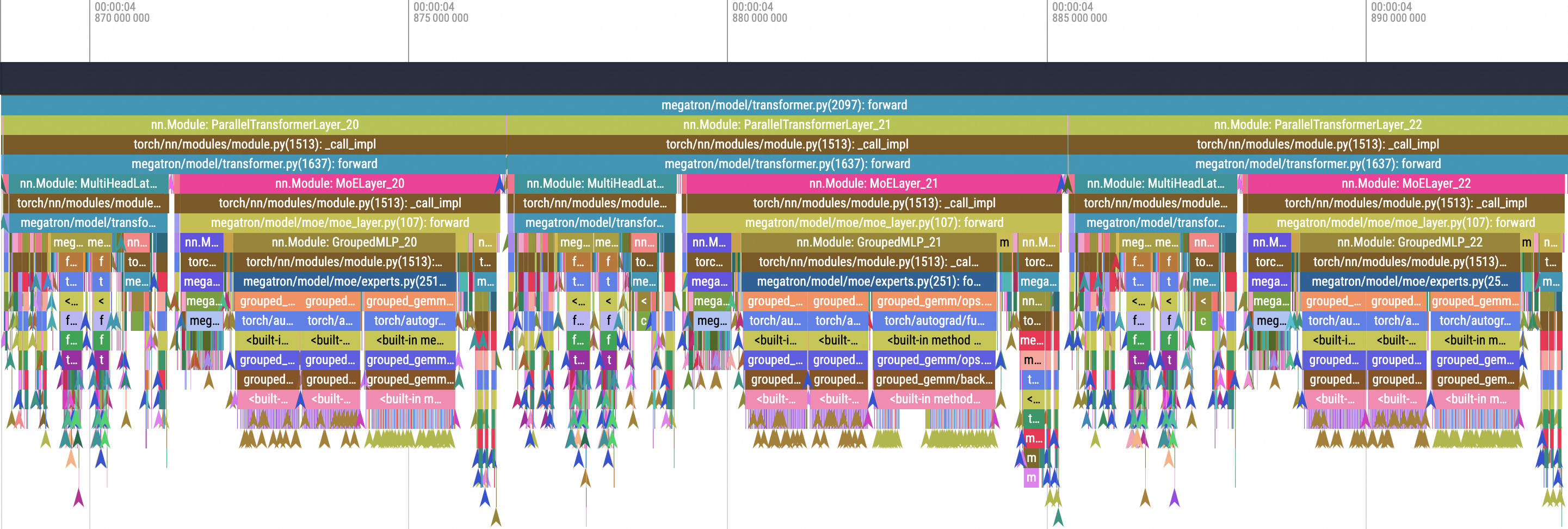}
    \caption{The python function behaviors of a forward step in the MoE training.}
    \label{fig:perfetto_moe_function}
\end{figure*}

\begin{figure*}[!t]
    \centering
    \includegraphics[width=\textwidth]{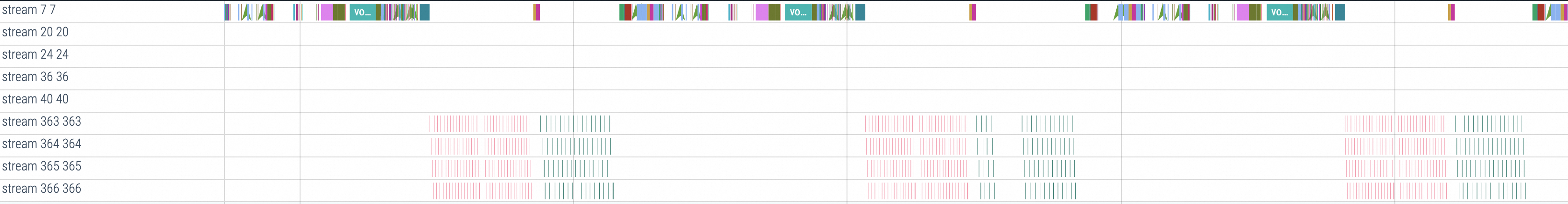}
    \caption{The kernel function (includes both computation and communication) behaviors of a forward step in the MoE training.}
    \label{fig:perfetto_moe_kernel}
\end{figure*}

\para{\sys 's Diagnosis result.}
\sys diagnosed both Version A and B, revealing that most GPU computation kernels and collective communications in Version B exhibited slightly higher $\beta$ values than Version A (Fig.~\ref{fig:case4_1} and \ref{fig:case4_2}). Since $\beta$ represents the time proportion within a single iteration and Version B had longer iteration durations, this indicates that most computation and communication functions consumed more time in Version B. In addition, \sys observed no significant difference in $\mu$ values between the two versions, which confirmed no hardware issues, indicating that Version B executed more workload on GPUs and networks during identical functions.

However, our customer confirmed that the workload of Version A and B were exactly the same.

\para{The root cause.} This \dlt job utilizes reinforcement learning techniques to train the data. It has both training processes and inference processes.
Inference processes should execute remotely but also persist in an idle state on the host due to developer oversight, running concurrently with training processes while repeatedly invoking \texttt{\small allgather} for synchronization.
Inference processes utilize gloo~\cite{gloo} to execute \texttt{\small allgather} by default, which is based on TCP, so they do not influence the training performance. However, in one commit (among the hundreds of commits between Version A and B), it was changed to NCCL (based on CUDA) to pursue better performance. The side-effect is that it contended for GPU resources with training processes (NCCL requires GPU SMs). This led to lower performance of both GPU computation and communication of training processes.

\para{Why \sys failed and how the root cause is identified.}
As discussed in \S\ref{limitations}, for most performance issues, when the problematic function and its behavior are identified, it is very close to the root cause, or we can even directly derive the root cause. In this case, however, there were too many "problematic" functions (including most GPU computation and communication functions). Moreover, no one was aware that training and inference are executed in the same host (in separate processes) in accident. Since only training processes have performance issues, we overlooked employing \sys to diagnose other processes (\eg, the idle inference process) on the same host.

The root cause is finally identified by human power. 
A team of 20 engineers performed a binary search among hundreds of commits between Version A and B, identifying a performance difference between two adjacent commits. This localization process took one month.

\para{Lessons and rethinking: Potential opportunities to automatically diagnose this case.}
As reinforcement learning becomes a promising technique of \dlt, some \dlt jobs execute processes of multiple actors in one host, like training and inference.

Based on the identified root cause, we retrospectively analyzed the diagnostic process. \sys had at least two opportunities to successfully locate this issue:
(1) We should first make an overview at all processes running on the host. Upon recognizing the potential concurrency between the (idle) inference and training processes in this \dlt, \sys should have been deployed to diagnose the idle inference process also, which would have directly revealed the NCCL function calls in the inference process.
(2) Even based solely on training process diagnostics, the root cause could have been indirectly deduced. \sys's detection of heavier workload execution in Version B versus Version A functions indicated resource contention.

The failure in problem localization primarily originated from our lack of an overview of the \dlt, causing us to overlook the impact of other processes (even existing accidentally) on the training procedure. For future performance diagnostics, especially for the \dlt with complex inter-process relationships, \sys will implement usability optimizations, such as automatically expanding the diagnosis scope to all \dlt-related processes within the host.

\section{Comparisons with State of the Arts}
\label{state-of-the-arts}

We compare \sys with recent monitoring techniques~\cite{jiang2024megascale,ncclprofiler,bpftrace} and profilers; Table~\ref{comparison} summarizes the results.

For online monitors, MegaScale~\cite{jiang2024megascale} deploys techniques to record CUDA-event timelines (to expose slow GPU kernels) and perform ms--s RDMA monitoring, but mainly report alerts/statistics; root-cause localization (especially for network issues) is typically manual. It also lacks Python function events, and thus cannot diagnose code-level issues. \camera{NCCL Profiler targets only communication by instrumenting the communication library. Bpftrace~\cite{bpftrace} selectively instruments key Python functions/kernels (\eg, via .so replacement) to detect slowdowns, but root-cause analysis still often requires manual effort.}
We also compare against directly using Nsight Systems~\cite{nsys} and Torch Profiler~\cite{torch_profiler}. Although they support offline/custom analysis (Nsight's Post-Collection Analysis API and Torch's JSON dumps), running them online incurs prohibitive overhead in production \dlt.

Table~\ref{comparison} further compares troubleshooting ability and diagnosis time on the problems in \S\ref{subsec-case1} and \S\ref{subsec-case2} (\eg, Case1-P1 represents Problem~1 of Case Study~1). Existing online monitors miss many issues due to incomplete data sources. Profiler traces cover more, but loading data from all workers can take days; since they do not provide built-in troubleshooting logic, we report only the data-loading time (end-to-end diagnosis would take longer).

\begin{table}[t]
\vspace{-7.5pt}
\caption{Comparisons with state-of-the-art approaches}
\vspace{-2.5pt}
\centering
\small
\setlength{\tabcolsep}{1.5pt}
\begin{tabular}{@{}ccccccccc@{}}
\toprule
\multirow{2}{*}{\makecell[l]{Technique}} 
& \multicolumn{3}{c}{Case 1} 
& \multicolumn{4}{c}{Case 2}  
& \multirow{2}{*}{\makecell[c]{Diagnostic time for\\10,000-GPU \dlt}} \\ 
\cmidrule(lr){2-4} \cmidrule(lr){5-8}  
& P1 & P2 & P3 & P1 & P2 & P3 & P4 & \\ \midrule  
MegaScale~\cite{jiang2024megascale} & \XSolidBrush & \XSolidBrush & \XSolidBrush & \XSolidBrush & \Checkmark & \XSolidBrush & \XSolidBrush & online \\
NCCL Profiler~\cite{ncclprofiler} & \XSolidBrush & \XSolidBrush & \XSolidBrush & \XSolidBrush & \Checkmark & \XSolidBrush & \XSolidBrush & online \\
bpftrace~\cite{bpftrace} & \Checkmark & \XSolidBrush & \Checkmark & \XSolidBrush & \XSolidBrush & \XSolidBrush & \XSolidBrush & online \\
Nsight System & \XSolidBrush & \XSolidBrush & \XSolidBrush & \Checkmark & \Checkmark & \XSolidBrush & \Checkmark & >1.5~days (offline)\\
Torch Profiler & \Checkmark & \Checkmark & \Checkmark & \XSolidBrush & \XSolidBrush & \Checkmark & \Checkmark & >3.5~days (offline)\\ 
\textbf{\sys} & \Checkmark & \Checkmark & \Checkmark & \Checkmark & \Checkmark & \Checkmark & \Checkmark & 3~min (online) \\ 
\bottomrule
\end{tabular}
\vspace{5pt}
\label{comparison}
\end{table}

\section{\edit{Profiling Overhead in Different Model Configurations}}
\label{overhead_with_config}
As explained in \S\ref{subsec-overhead}, with \sys, profiling and data generation are the two components executed in the main process of \dlt (Figure~\ref{fig:overload}),
Table~\ref{tab:overload} evaluates them systematically in different \dlt configurations.
In general, the time spent on data generation correlates positively with the number of function executions on a single worker within the profiling window. Increasing the degree of parallelism (\eg, the tensor parallelism, TP) results in more fragmented model training. This fragmentation leads to more function execution events, increasing the data generation time.

\begin{table}[!t]
\vspace{-10pt}
\caption{Overhead in different model configurations}
\centering
\small
\begin{tabular}{@{}cccccc@{}}
\toprule
\multicolumn{3}{c}{Configurations} & \multicolumn{3}{c}{Overhead} \\
\hline
\multirow{2}{*}{model size}&\multirow{2}{*}{tp}&\multirow{2}{*}{pp}&training&profiling&generate data \\
&&&(s/iter)&(s/iter)&(s)\\
\midrule
\multirow{2}{*}{gpt3-7b}
&1&1&1.371&1.389&13 \\
&2&1&1.777&2.002(+12\%)&19 \\
\hline
\multirow{3}{*}{gpt3-13b}
&2&1&2.489&2.485&17 \\
&4&1&1.228&1.425(+16\%)& 26 \\
&8&1&1.528&1.707(+11\%)&28 \\
\hline
\multirow{2}{*}{gpt3-65b} 
& 8&4&1.191&1.202&15 \\
& 8&8 &1.281&1.288&10 \\
\bottomrule
\end{tabular}
\vspace{10pt}
\label{tab:overload}
\end{table}

In most configurations, profiling does not introduce overhead to training. However, when the model is small and the parallelism parameters are large (\eg, GPT-3 7B with TP=2; GPT-3 13B with TP=4 or 8), there is a major increase in training time. This is because such configurations impose a higher load on the CPU; as profiling requires CPU computation, it causes contention.
Fortunately, these impractical configurations 
are never used in production. Note: \sys is still effective in these configurations and the overhead only happens during the profiling window. 

\section{Examples of \dlt executions}
\label{appendix-behaviors}

Figure~\ref{fig:perfetto_moe} shows the timeline (using \href{https://ui.perfetto.dev/}{https://ui.perfetto.dev/}) of high-level Python function call stacks of two adjacent training iterations of one of our production MoE models. Each iteration includes multiple forward and backward phases, and their function runtime behaviors are identical. All workers are executing the same functions with the same behaviors. Due to space limitations, we only present one of them.

Figure~\ref{fig:perfetto_moe_function} and \ref{fig:perfetto_moe_kernel} zoom into one of the forward phase in Figure~\ref{fig:perfetto_moe}. It shows that in the forward phase, low-level functions are also executed repeatedly, and their behaviors (\eg, duration of execution) are highly similar across multiple executions.

\camera{In large model training, especially in heterogeneous computation, workers may not exhibit identical execution behavior due to differences in parallelization roles (\eg, pipeline/expert parallelism), input variability (\eg, variable-length video samples), and system effects (\eg, scheduling and contention). Nevertheless, worker-side execution is usually structured as repeated iterations that invoke a largely stable set of training-step functions and GPU kernels. Moreover, tensor shapes are often fixed by model configuration and batching policy, or are restricted to a small set of discrete shapes through padding, bucketing, truncation, or shape-constrained micro-batching. As a result, per-function runtime and resource-usage metrics across workers are expected to be either broadly consistent or, when not identical, to follow a relatively stable distribution reflecting systematic role differences and stochastic runtime noise. Under this assumption, function instances that fall outside the expected cross-worker distribution—\ie, statistically significant outliers in runtime or resource consumption—can be treated as anomalous behavior.}

\end{document}